\documentclass[11pt,a4paper]{article}

\usepackage{jheppub}
\usepackage{latexsym}
\usepackage{multirow}
\usepackage{color}
\usepackage[usenames,dvipsnames,table]{xcolor}
\usepackage{graphicx}
\usepackage{epsfig}  
\usepackage{epsf}    
\usepackage{dcolumn}
\usepackage{bm}
\usepackage{dcolumn}
\usepackage{textcomp}
\usepackage{float}
\usepackage{subfig}
\usepackage{hypcap}
\usepackage[]{hyperref}
\usepackage{makecell}
\usepackage{color}
\usepackage{pifont}
\usepackage{appendix}
\usepackage{amsmath}
\usepackage{multirow,bigdelim}  
\usepackage{lineno}
\usepackage[normalem]{ulem}
\hypersetup{
  bookmarks=true,         
  unicode=false,          
  pdftoolbar=true,        
 pdfmenubar=true,        
 pdffitwindow=true,     
 pdfstartview={FitH},    
 pdfsubject={Neutrino Oscillations Phenomenology},   
 pdfnewwindow=true,      
 pdfcreator={RevTeX},
 colorlinks=true,       
 linkcolor=red,          
 citecolor=blue,        
 filecolor=black,      
 urlcolor=blue,           
  }

\newcommand{\be}{\begin{equation}}
\newcommand{\ee}{\end{equation}}
\newcommand{\ba}{\begin{eqnarray}}
\newcommand{\ea}{\end{eqnarray}}

\newcommand{\capdef}{}
\newcommand{\mycaption}[2][\capdef]{\renewcommand{\capdef}{#2}
       \caption[#1]{{\footnotesize #2}}}
\makeatletter
\renewcommand{\fnum@table}{\textbf{\tablename~\thetable}}
\renewcommand{\fnum@figure}{\textbf{\figurename~\thefigure}}
\makeatother

\preprint{IP/BBSR/2021-2}

\title{Validating the Earth's Core using Atmospheric Neutrinos with ICAL at INO}

\author[a,b,c]{Anil Kumar,}
\author[a,c,d]{Sanjib Kumar Agarwalla}

\affiliation[a]{Institute of Physics, Sachivalaya Marg, Sainik School Post,
  Bhubaneswar 751005, India}
\affiliation[b]{Applied Nuclear Physics Division, Saha Institute of
  Nuclear Physics, Block AF, Sector 1, Bidhannagar, Kolkata 700064, India}
\affiliation[c]{Homi Bhabha National Institute, Anushakti Nagar,
  Mumbai 400094, India}
\affiliation[d]{International Centre for Theoretical Physics,
  Strada Costiera 11, 34151 Trieste, Italy}

\emailAdd{anil.k@iopb.res.in (ORCID: 0000-0002-8367-8401)}
\emailAdd{sanjib@iopb.res.in (ORCID: 0000-0002-9714-8866)}

\abstract
{
	The Iron Calorimeter (ICAL) detector at the proposed India-based Neutrino Observatory (INO) aims to detect atmospheric neutrinos and antineutrinos separately in the multi-GeV range of energies and over a wide range of baselines. By utilizing its charge identification capability, ICAL can efficiently distinguish $\mu^-$ and $\mu^+$ events. Atmospheric neutrinos passing long distances through Earth can be detected at ICAL with good resolution in energy and direction, which enables ICAL to see the density-dependent matter oscillations experienced by upward-going neutrinos in the multi-GeV range of energies. In this work, we explore the possibility of utilizing neutrino oscillations in the presence of matter to extract information about the internal structure of Earth complementary to seismic studies. Using good directional resolution, ICAL would be able to observe 331 $\mu^-$ and 146 $\mu^+$ core-passing events with 500 kt$\cdot$yr exposure. With this exposure, we show for the first time that the presence of Earth's core can be independently confirmed at ICAL with a median $\Delta \chi^2$ of 7.45 (4.83) assuming normal (inverted) mass ordering by ruling out the simple two-layered mantle-crust profile in theory while generating the prospective data with the PREM profile. We observe that in the absence of charge identification capability of ICAL, this sensitivity deteriorates significantly to 3.76 (1.59) for normal (inverted) mass ordering. 
}

\arxivnumber{2104.11740}

\begin{document}
\maketitle
\flushbottom

\section{Introduction and motivation}
\label{sec:introduction}

Neutrinos are elusive particles, but they are capable of reaching places inaccessible by any other means. The tiny interaction cross section enables neutrinos to even pass through solid objects like Earth because they only interact via weak interactions. Neutrinos undergo flavor change as they move in space and time. This phenomenon is known as neutrino flavor oscillation~\cite{Pontecorvo:1967fh}. The Super-Kamiokande (Super-K) was the first experiment to discover neutrino oscillation using atmospheric neutrino data in 1998~\cite{Fukuda:1998mi}. The atmospheric neutrinos are produced during the interaction of cosmic rays with the atmosphere, and they travel long distances through the Earth. The atmospheric neutrinos undergo coherent elastic forward scattering with electrons inside the Earth which leads to the modification of neutrino oscillations. When neutrinos pass deep through the mantle, the Mikheyev-Smirnov-Wolfenstein (MSW) resonance~\cite{Wolfenstein:1977ue,Mikheev:1986gs,Mikheev:1986wj} starts playing an important role in neutrino oscillations around 6 to 10 GeV of energies. 
On the other hand, the core-passing neutrinos with energies in the range of 3 to 6 GeV experience a different kind of resonant effect which is known as neutrino oscillation length resonance (NOLR)~\cite{Petcov:1998su,Chizhov:1998ug,Petcov:1998sg,Chizhov:1999az,Chizhov:1999he} or parametric resonance~\cite{Akhmedov:1998ui,Akhmedov:1998xq}. These density-dependent matter effects can be used to reveal the distribution of matter inside the Earth. 

The neutrinos have the potential to throw some light on the internal structure of Earth via neutrino absorption, oscillations, and diffraction. The idea of exploring Earth's interior using neutrino absorption dates back to 1974~\cite{Volkova1974} where attenuation of neutrino is exploited at energies greater than 10 TeV~\cite{Gandhi:1995tf}. There are numerous studies considering neutrinos from different sources, such as man-made neutrinos~\cite{Volkova1974,Nedyalkov:1981,Nedyalkov:1981yy,1983BlDok..36.1515N,DeRujula:1983ya,Wilson:1983an,Askarian:1985ca,Volkova:1985zc, Tsarev:1986ay,Borisov:1986sm,Tsarev:1986xg}, extraterrestrial neutrinos~\cite{Wilson:1983an,Kuo:1995,Crawford:1995,Jain:1999kp,Reynoso:2004dt} and atmospheric neutrinos~\cite{GonzalezGarcia:2007gg,Borriello:2009ad,Takeuchi2010,Romero:2011zzb}. The neutrino-based absorption tomography of Earth has been performed using atmospheric neutrino data at IceCube detector~\cite{Donini:2018tsg}. On the other hand, the neutrino oscillation tomography relies on the matter effects in neutrino oscillations which has been considered by the study of man-made beams~\cite{Ermilova:1986ph,Nicolaidis:1987fe,Ermilova:1988pw,Nicolaidis:1990jm,Ohlsson:2001ck,Ohlsson:2001fy,Winter:2005we,Minakata:2006am,Gandhi:2006gu,Tang:2011wn,Arguelles:2012nw}, atmospheric~\cite{Agarwalla:2012uj,Rott2015,Winter:2015zwx,Bourret2017}, solar~\cite{Ioannisian:2002yj,Akhmedov2005}, and supernova~\cite{Akhmedov2005,LINDNER2003755} neutrinos. The third possibility of Earth tomography using the study of diffraction pattern produced by coherent neutrino scattering in crystalline matter inside Earth is technologically not feasible~\cite{Fortes2006}.

The current understanding of the structure of Earth is provided by seismic studies~\cite{Robertson:1966,Dziewonski:1981xy,Loper:1995,Alfe:2007} where the propagation of seismic waves inside the Earth reveals the properties of matter. The Earth consists of concentric shells of different densities and compositions. The outermost surface of Earth is made up of solid crust, below which we have a viscous mantle made up of silicate oxide. The mantle is followed by a high-density core of iron-alloy. The information carried by the seismic waves may get altered on its way. On the contrary, the information on the interaction of neutrinos with ambient electrons (so-called Earth matter effect) remains unaltered when neutrinos travel long distances inside Earth. But owing to its weak interaction nature, usually event rates are not very large in neutrino experiments, and we need to compensate for it by using massive detectors and large exposures. A neutrino detector with good resolution in the multi-GeV range of energy and direction of neutrino will be able to observe modified event distribution due to neutrino oscillations in the presence of matter.  

The Iron Calorimeter (ICAL) detector at the proposed India-based Neutrino Observatory (INO)~\cite{Kumar:2017sdq} would be able to detect neutrinos and antineutrinos separately in the multi-GeV range of energies covering baselines over a wide range of 10 to $10^4$ km. Due to the presence of a magnetic field of 1.5 Tesla~\cite{Behera:2014zca}, ICAL would be able to distinguish between $\mu^-$ and $\mu^+$ events separately. The ICAL has a very high resolution of direction and energy for upward-going muons in the energy range of 1 to 10 GeV, which enables ICAL to observe the matter effect felt by neutrinos. Exploring Earth matter effect separately in neutrino (by observing $\mu^-$ events) and antineutrino (by observing $\mu^+$ events) modes through their mass-induced flavor oscillations inside the Earth plays an important role to probe the inner structure of Earth, which we demonstrate explicitly while presenting our main results later. The MSW resonance can be observed around 6 to 10 GeV of energies and provides crucial information about the mantle. On the other hand, vertically upward-going neutrinos with large baselines pass through the high-density core and feel the NOLR/parametric resonance around 3 to 6 GeV of energies. In this work, we will study the impact of the presence of various layers inside Earth on neutrino oscillations and perform statistical analysis to establish the presence of a high-density core inside Earth by ruling out the mantle-crust profile with respect to (w.r.t.) the core-mantle-crust profile.

In Section~\ref{sec:earth_model}, we discuss the internal structure of Earth known from seismic studies and describe the profiles of Earth to be probed by neutrino oscillations in this work. The oscillation probabilities in the presence of matter governed by various profiles of Earth are described in Section~\ref{sec:probability}. Next, we explain the method to simulate neutrino events at ICAL in Section~\ref{sec:events}. The good directional resolution at ICAL is used to identify neutrinos passing through core, mantle, and crust in Section~\ref{sec:events_layer} which also describes the resultant distribution of reconstructed muon events for these neutrinos passing through a particular set of layers. The method for statistical analysis is explained in Section~\ref{sec:statistical analysis} which is followed by the results in terms of the statistical significance for establishing core and ruling out alternative profiles of Earth in Section~\ref{sec:results}. Finally, we conclude in Section~\ref{sec:conclusion}.

\section{A brief review of the internal structure of Earth}
\label{sec:earth_model}

\begin{figure}[htb!]
	\centering
	\includegraphics[width=0.4\textwidth]{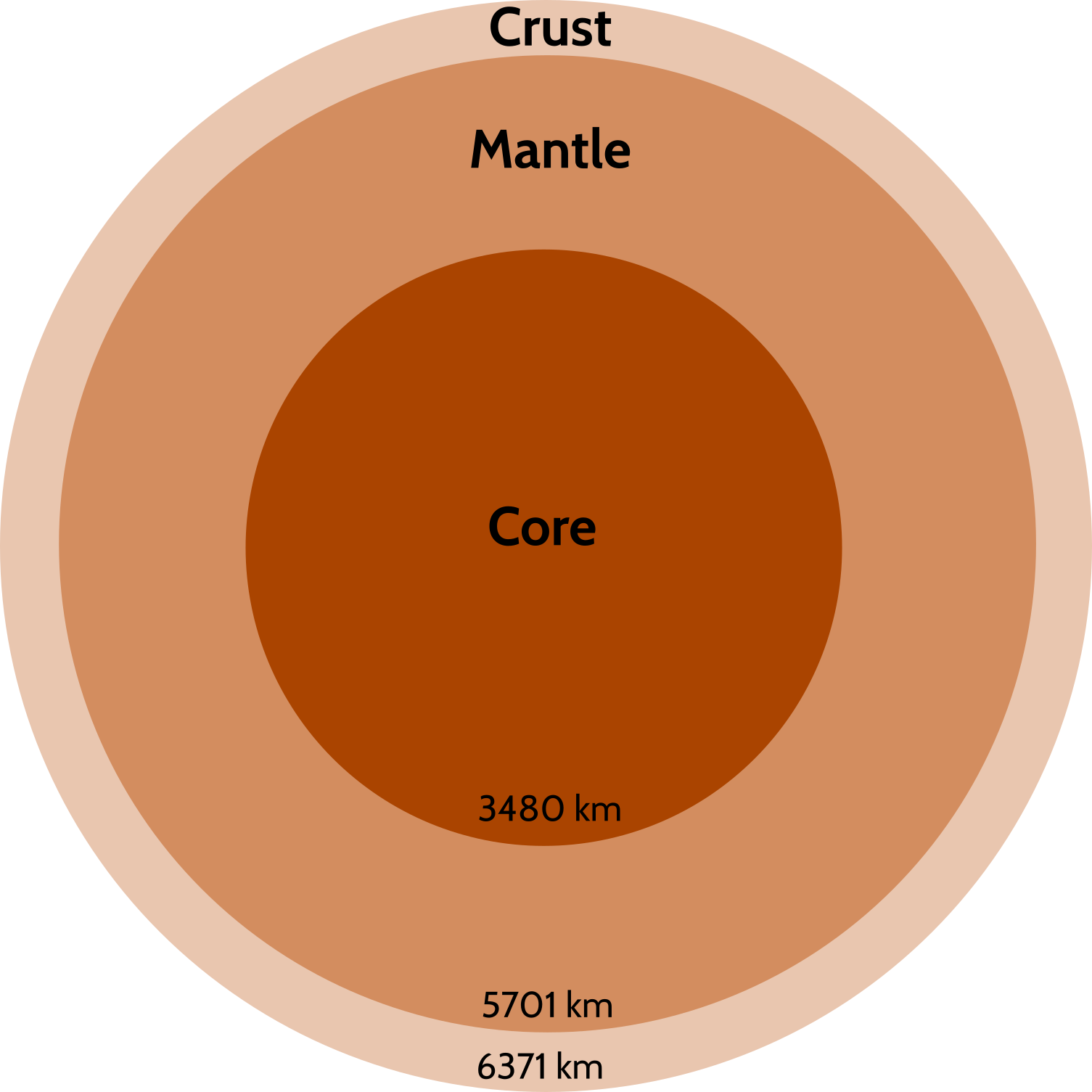}
	\includegraphics[width=0.5\textwidth]{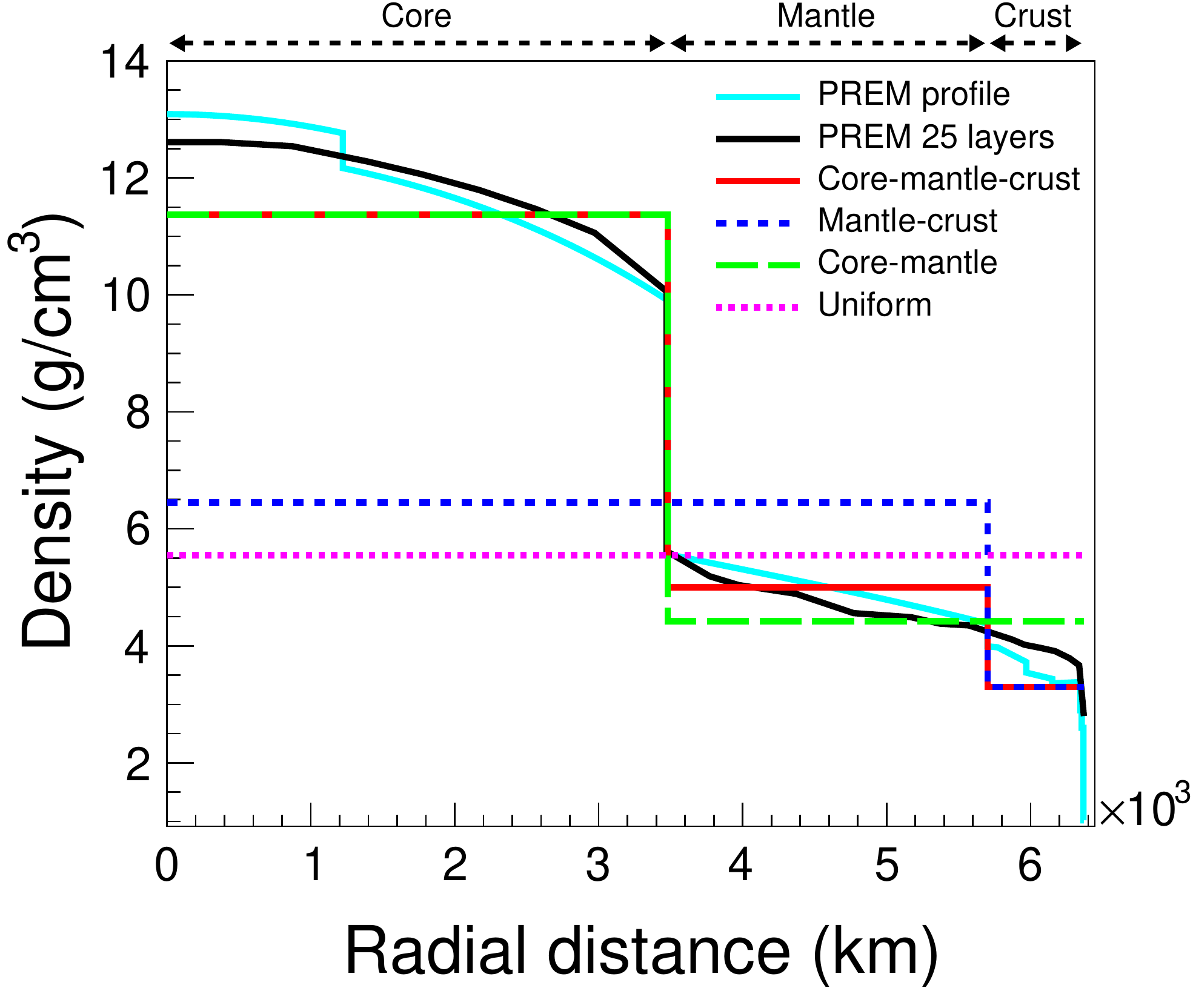}
	\mycaption{Left: Three-layered profile of Earth. Right: Density distribution of profiles of Earth as a function of radial distance from the center of Earth. Note that the total mass of the Earth is the same in all the profiles.}
	\label{fig:three-layer-model}
\end{figure}

The seismic studies have revealed that Earth consists of concentric shells, which are crust, mantle, and core, each of them is further divided into subshells with different properties~\cite{Robertson:1966,Loper:1995, Alfe:2007}. The crust constitutes about 0.4\% mass of Earth, whereas the mantle and core contributions are about 68\% and 32\%, respectively. The radius of the core is almost half the radius of Earth, whereas the density of the core is twice that of the mantle. 

The outermost layer crust is made up of solid rocks and has the lowest density among all layers~\cite{Robertson:1966,Alfe:2007}. Under the crust, we have the mantle, which consists of extremely hot rocks that are solid in the upper mantle but highly viscous plastic in the lower mantle. The mantle is followed by the high-density core, which is mainly composed of iron and nickel. The core can further be divided into outer core and inner core. The shear (S) waves are unable to transmit through the outer core, whereas the velocity of compressional (P) waves decreases significantly. This observation indicates that the outer core is composed of fluid with viscosity as low as that of water. The inner core is made up of solid metal because it allows the propagation of both S and P waves.

The detailed distribution of density inside Earth is available in the Preliminary Reference Earth Model (PREM)~\cite{Dziewonski:1981xy} as shown by the cyan curve in the right panel of Fig.~\ref{fig:three-layer-model} where the density is shown as a function of radial distance \textit{i.e.} the distance of a layer from the center of Earth. We would like to mention that in the actual PREM profile, the Earth is divided into 81 layers. But what we use here as a PREM profile~\cite{Dziewonski:1981xy} for the sake of computational ease is a 25-layered profile of Earth (black curve) that preserves all the important features of the Earth profile. We have checked that whatever conclusion, we have drawn in this paper, will not alter whether we take 25 layers or 81 layers.

Guided by the PREM profile of Earth, we consider a three-layered profile of Earth as shown in the left panel of  Fig.~\ref{fig:three-layer-model}. The innermost layer is the core which is followed by the mantle, and the outermost layer is the crust. The density distribution for the three-layered structure is shown by the red curve in the right panel of Fig.~\ref{fig:three-layer-model}. The layer boundaries and their densities for the three-layered profile of Earth are mentioned in Table~\ref{tab:profiles}.

\begin{table}[htb!]
	\centering
	\begin{tabular}{|c|c|c|}
		\hline \hline
		Profiles & Layer boundaries (km) & Layer densities (g/cm\textsuperscript{3})\\
		\hline
		PREM & 25 layers & 25 densities\\
		Core-mantle-crust & (0, 3480, 5701, 6371) & (11.37, 5, 3.3) \\
		Mantle-crust  & (0, 5701, 6371) & (6.45, 3.3)\\
		Core-mantle  & (0, 3480, 6371) & (11.37, 4.42)\\
		Uniform  & (0, 6371) & (5.55) \\
		\hline \hline
	\end{tabular}\\
	\mycaption{The boundaries and densities of layers for various profiles of Earth considered in this analysis. The radius and mass of Earth remain invariant for all these profiles.}
	\label{tab:profiles}
\end{table}

Since neutrino oscillations occur in vacuum also, so one of the important tasks is to rule out the vacuum hypothesis and feel the presence of matter. For vacuum, we consider the density of the Earth to be zero. We further consider alternative profiles of the Earth as mentioned in Table~\ref{tab:profiles} for testing against the three-layered profile using neutrino oscillations. While considering alternative profiles of Earth, we assume the radius and the mass of Earth to be invariant\footnote{Note that the moment of inertia of Earth can also be considered as an additional invariant quantity on which the information is obtained from gravitational studies independent of seismology.}. The dashed blue curve in the right panel of Fig.~\ref{fig:three-layer-model} shows the mantle-crust profile, which has a two-layered structure with mantle and crust where the core and mantle are fused together. The core-mantle profile has a two-layered structure with core and mantle where the crust is merged into the mantle as shown by the dashed green curve in the right panel of Fig.~\ref{fig:three-layer-model}. The uniform density profile is shown by the dotted pink curve in the right panel of Fig.~\ref{fig:three-layer-model}. 

Since the distributions of densities in these profiles of the Earth are different from each other, we expect the neutrino oscillation probability to modify differently in the presence of matter governed by these profiles. In Section~\ref{sec:probability}, we discuss the effect of these profiles of the Earth on neutrino oscillation probabilities.

\section{Effect of various density profiles of Earth on oscillograms}
\label{sec:probability}

The interactions of cosmic rays with nuclei of the atmosphere produce unstable charged particles like pions and kaons whose decay chains result in both muon and electron type of neutrinos as well as antineutrinos. The ratio of total neutrinos and antineutrinos of muon type with that of electron type is approximately 2. ICAL is sensitive to muon neutrinos and antineutrinos in the multi-GeV range of energy. After traveling long distance inside the Earth, the initial muon neutrino $\nu_\mu$ at production may survive as muon neutrino $\nu_\mu$ at detection with survival probability $P(\nu_\mu \rightarrow \nu_\mu)$ whereas an electron neutrino $\nu_e$ may oscillate to muon neutrino $\nu_\mu$ with appearance probability $P(\nu_e \rightarrow \nu_\mu)$. The muon neutrino events detected at ICAL are contributed by both survival $(\nu_\mu \rightarrow \nu_\mu)$ as well as appearance $(\nu_e \rightarrow \nu_\mu)$ channels. 

\begin{table}[htb!]
	\centering
	\begin{tabular}{|c|c|c|c|c|c|c|}
		\hline
		$\sin^2 2\theta_{12}$ & $\sin^2\theta_{23}$ & $\sin^2 2\theta_{13}$ & $\Delta m^2_
		\text{eff}$ (eV$^2$) & $\Delta m^2_{21}$ (eV$^2$) & $\delta_{\rm CP}$ & Mass Ordering\\
		\hline
		0.855 & 0.5 & 0.0875 & $2.49\times 10^{-3}$ & $7.4\times10^{-5}$ & 0 & Normal (NO)\\
		\hline 
	\end{tabular}
	\mycaption{The benchmark values of neutrino oscillation parameters used in this analysis. These values are consistent with the present neutrino global fits~\cite{Marrone:2021,NuFIT,Esteban:2020cvm,deSalas:2020pgw}. 
    Normal mass ordering indicates $m_1 < m_2 < m_3$.}
	\label{tab:osc-param-value}
\end{table}

\begin{figure}[htb!]
	\centering
	\includegraphics[width=0.48\linewidth]{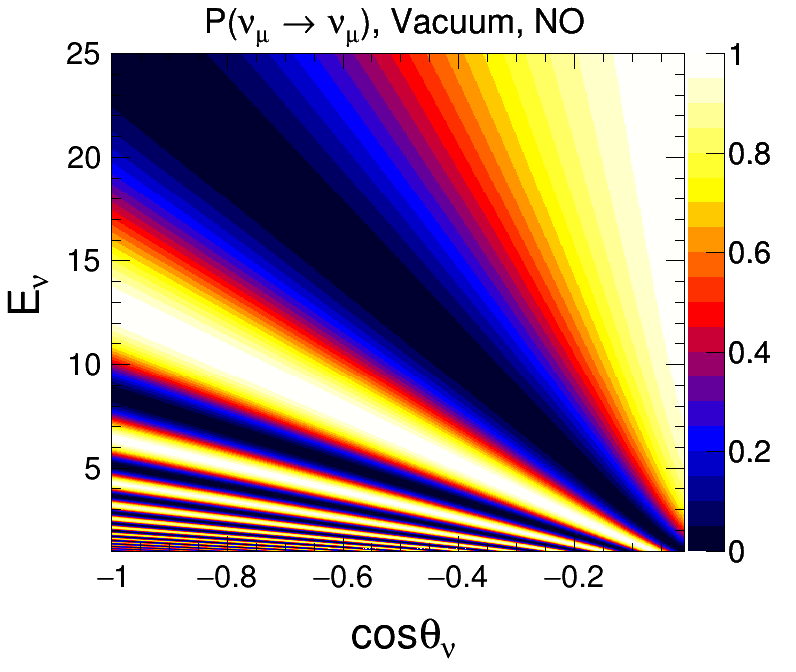}
	\includegraphics[width=0.48\linewidth]{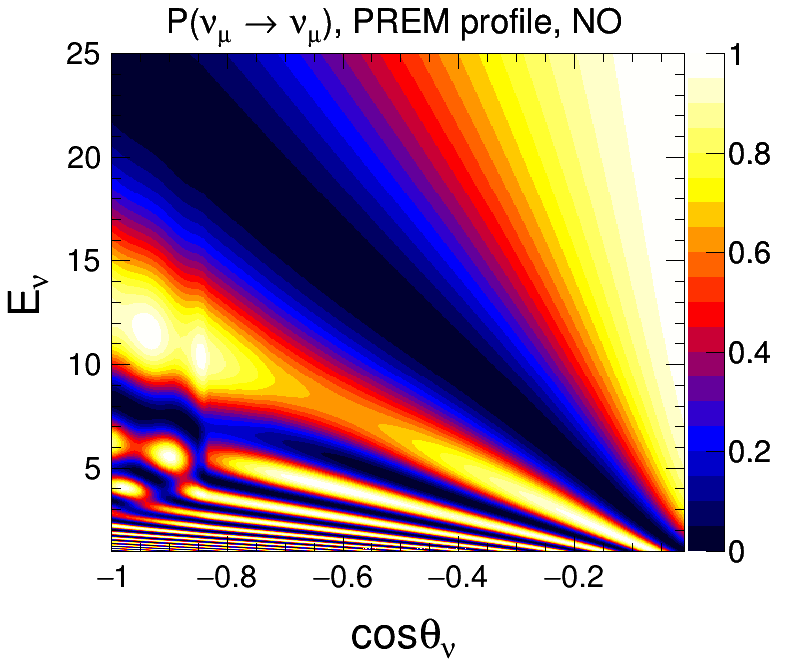}
	\includegraphics[width=0.48\linewidth]{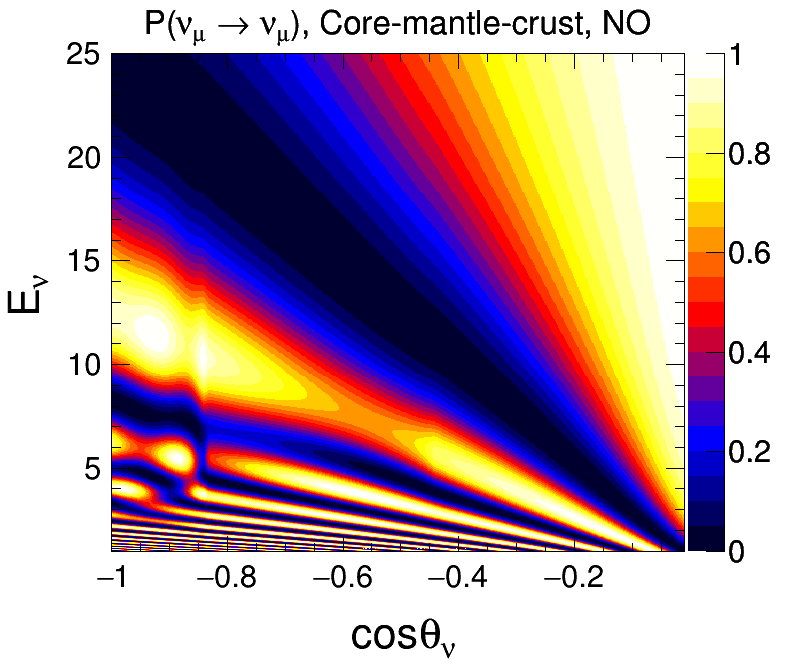}
	\includegraphics[width=0.48\linewidth]{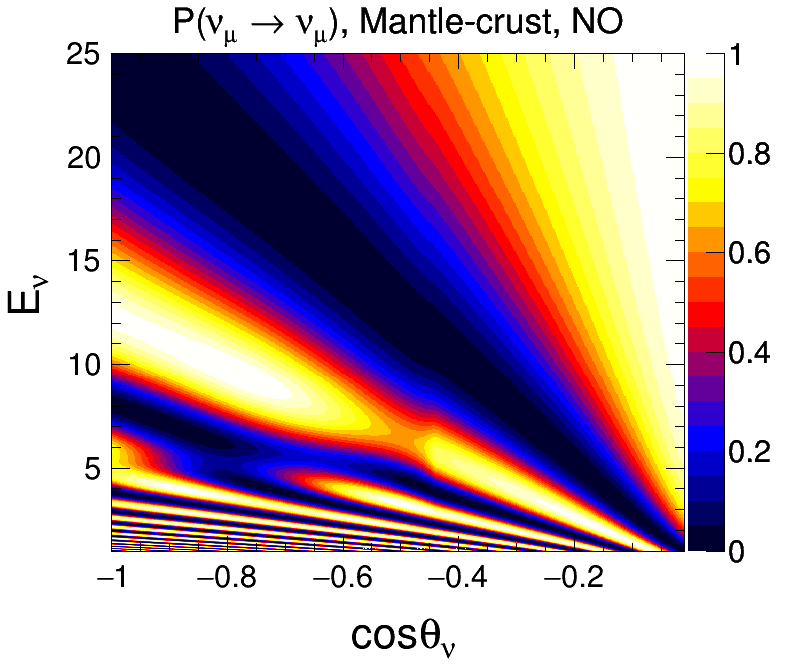}
	\includegraphics[width=0.48\linewidth]{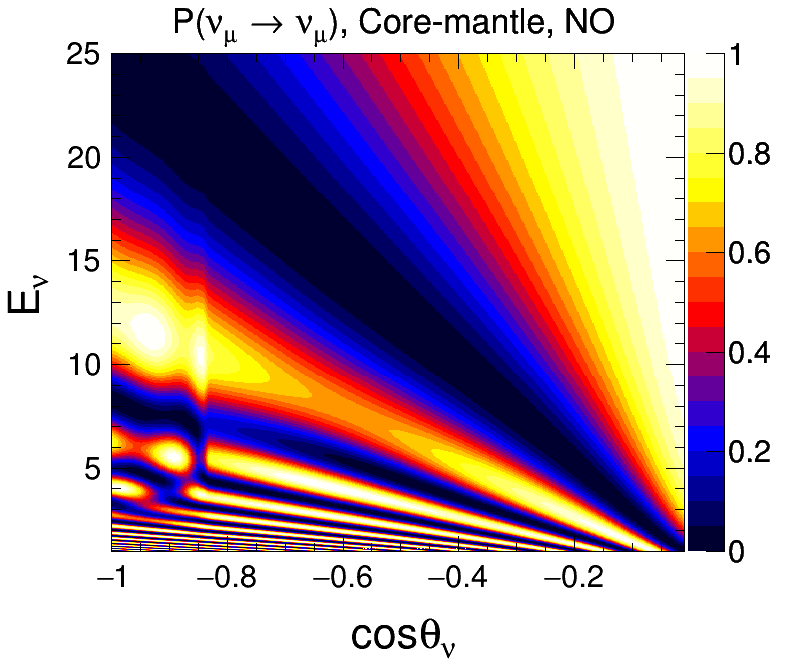}
	\includegraphics[width=0.48\linewidth]{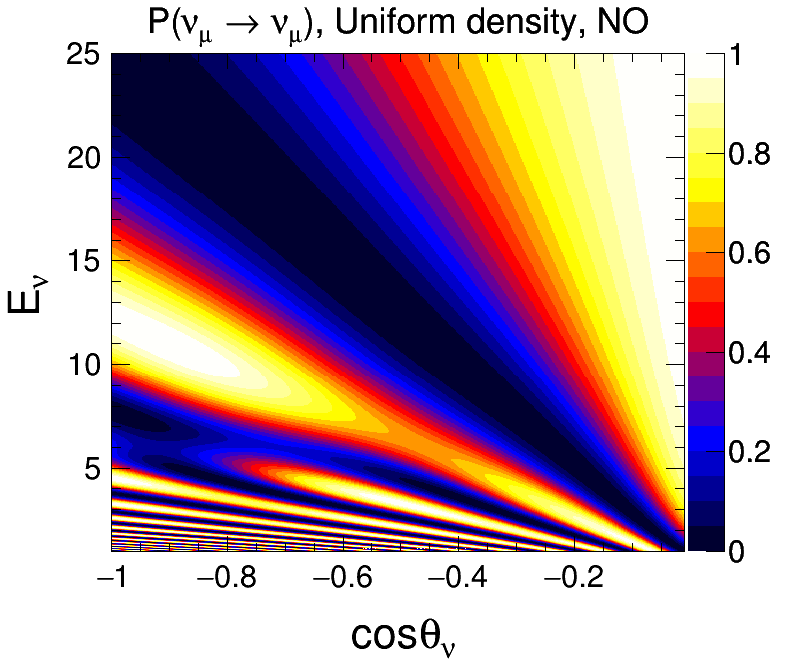}
	\mycaption{$P(\nu_{\mu} \rightarrow \nu_{\mu})$ oscillograms considering various density profiles of Earth.
	We take the three-flavor oscillation parameters from Table~\ref{tab:osc-param-value}. We assume NO and $\sin^2\theta_{23} = 0.5$.}
	\label{fig:Puu_oscillogram_model}
\end{figure}

\begin{figure}[htb!]
	\centering
	\includegraphics[width=0.48\linewidth]{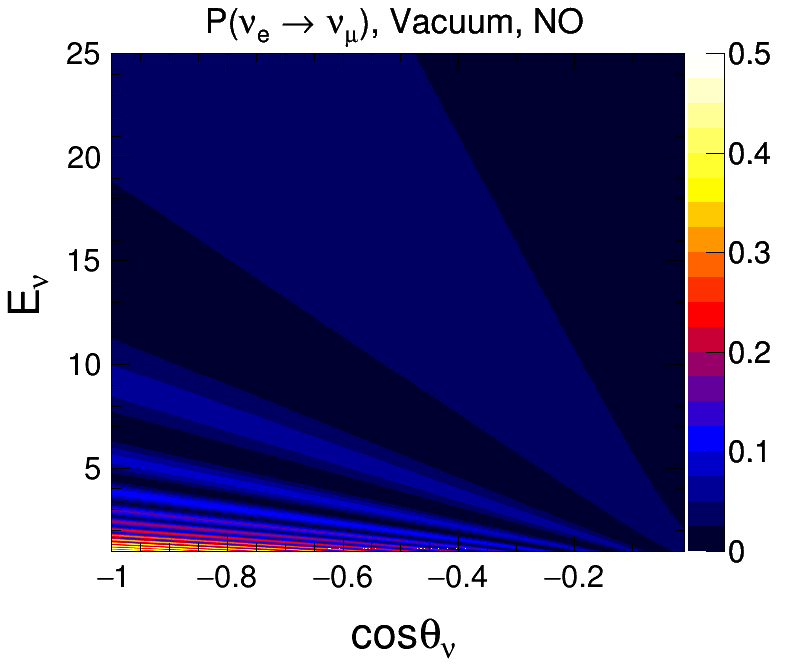}
	\includegraphics[width=0.48\linewidth]{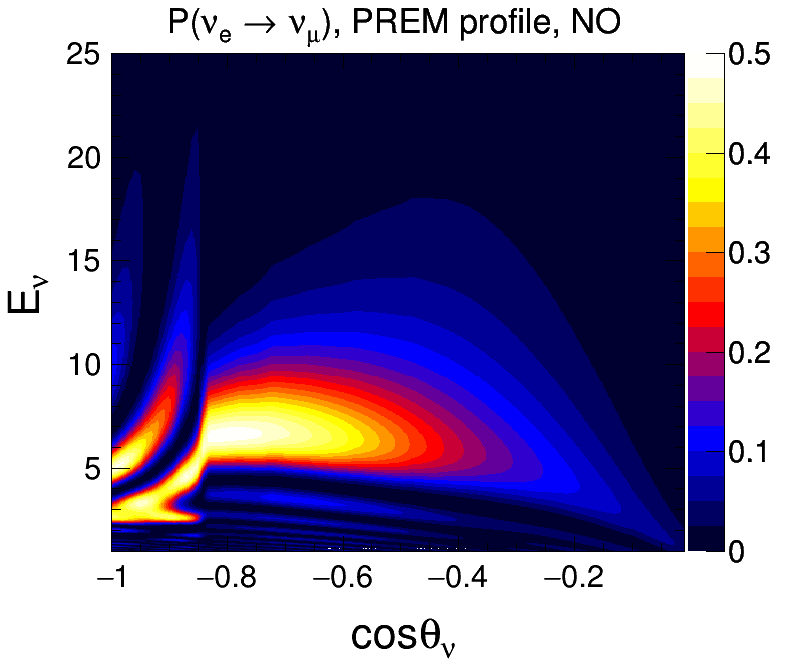}
	\includegraphics[width=0.48\linewidth]{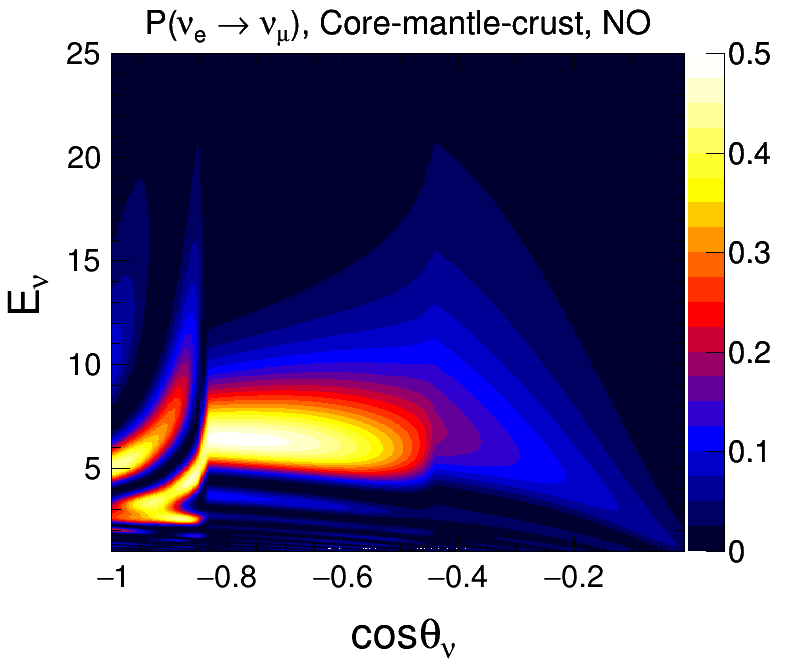}
	\includegraphics[width=0.48\linewidth]{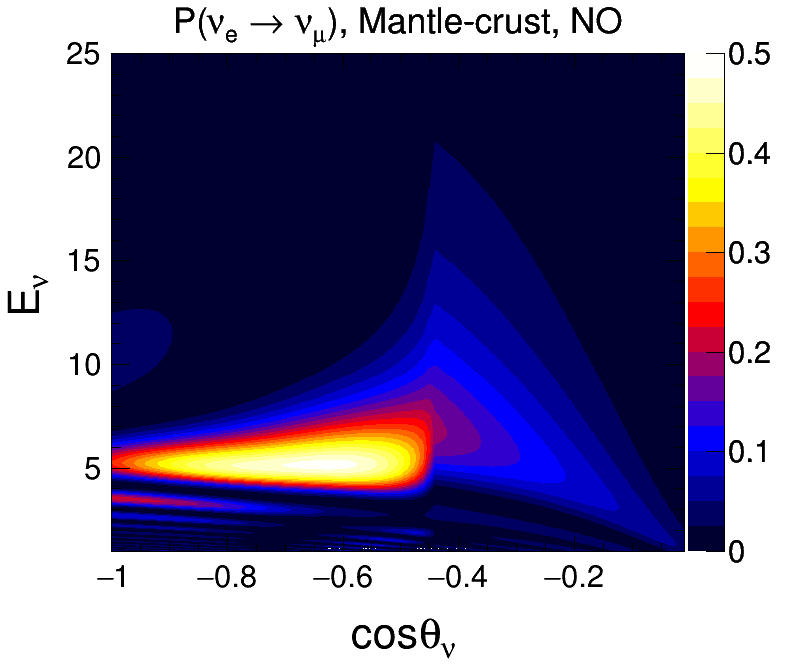}
	\includegraphics[width=0.48\linewidth]{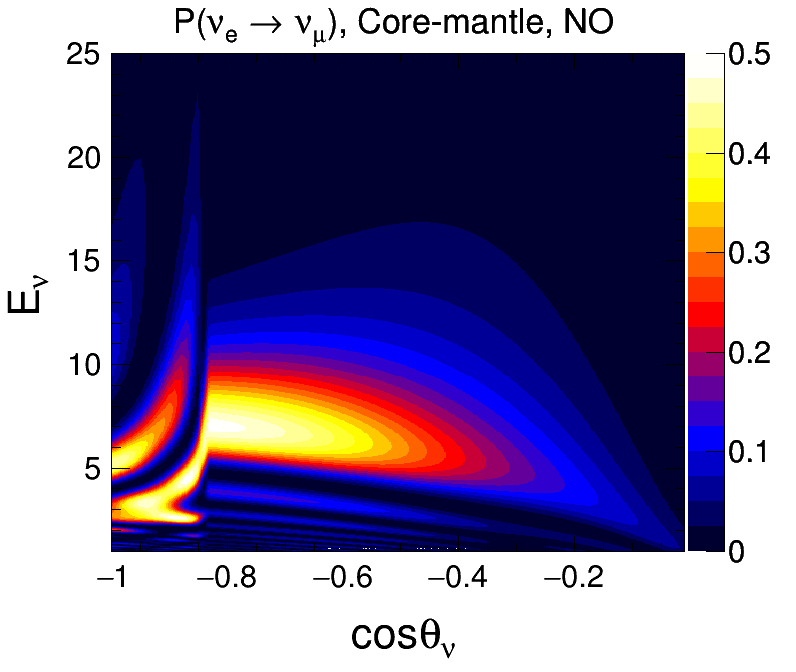}
	\includegraphics[width=0.48\linewidth]{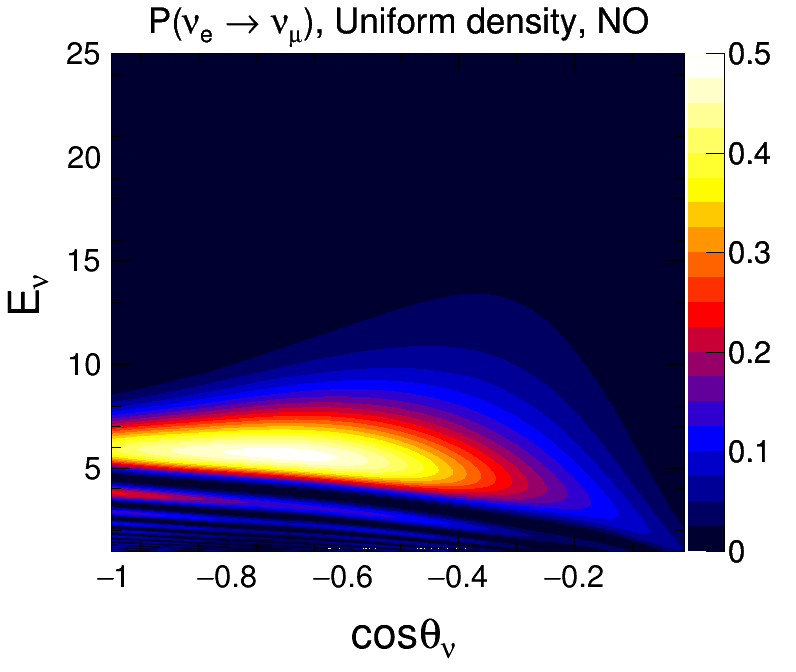}
	\mycaption{$P(\nu_{e} \rightarrow \nu_{\mu})$ oscillograms considering various density profiles of Earth.
    We take the three-flavor oscillation parameters from Table~\ref{tab:osc-param-value}. We assume NO and $\sin^2\theta_{23} = 0.5$.}
	\label{fig:Peu_oscillogram_model}
\end{figure}

In this analysis, we use the values of benchmark oscillation parameters mentioned in Table~\ref{tab:osc-param-value}. We use the effective atmospheric mass-squared difference\footnote{The effective atmospheric mass-squared difference is related to $\Delta m^2_{31}$ and $\Delta m^2_{21}$ as follows~\cite{deGouvea:2005hk,Nunokawa:2005nx}
\begin{equation}
\Delta m^2_\text{eff} = \Delta m^2_{31} - \Delta m^2_{21} (\cos^2\theta_{12} - \cos \delta_\text{CP} \sin\theta_{13}\sin2\theta_{12}\tan\theta_{23}).
\end{equation}} $\Delta m^2_\text{eff}$ to consider mass ordering (MO), the positive and negative value of $\Delta m^2_\text{eff}$ corresponds to normal ordering (NO, $m_1 < m_2 < m_3$) and inverted ordering (IO, $m_3 < m_1 < m_2$), respectively. The standard $W$-mediated matter potential $V_{CC}$ experienced by neutrino/antineutrino during interaction with the ambient electrons in the matter can be expressed as 
\begin{equation}\label{eq:matter_pot}
V_{CC} = \pm \sqrt2 G_F N_e \approx \pm7.6 \times Y_e \times 10^{-14} \left[\frac{\rho}{\text{g/cm}^3}\right]~\text{eV}\, ,
\end{equation}
where, $Y_e = N_e/(N_p +N_n)$ corresponds to the relative electron number density inside the matter and $\rho$ denotes the matter density of various layers inside the Earth for a given profile. The positive (negative) sign is for neutrino (antineutrino). In the present analysis, we assume the Earth to be electrically neutral and isoscalar where $N_n \approx N_p = N_e$ which results in $Y_e = 0.5$. 
In Fig.~\ref{fig:Puu_oscillogram_model}, we present the oscillograms\footnote{In Ref.~\cite{Akhmedov:2006hb}, the authors gave a detailed physics interpretation of the oscillograms in terms of the amplitude and phase conditions while describing various features such as MSW peaks, parametric ridges, local maxima, zeros, and saddle points.} for $\nu_\mu$ survival channel $(\nu_\mu \rightarrow \nu_\mu)$ in the plane of $\cos\theta_\nu$ vs. $E_\nu$ considering NO in the three-flavor neutrino oscillation framework using various profiles of Earth. In Fig.~\ref{fig:Peu_oscillogram_model}, we show the same for $\nu_\mu$ appearance channel $(\nu_e \rightarrow \nu_\mu)$.   
In Figs.~\ref{fig:Puu_oscillogram_model} and~\ref{fig:Peu_oscillogram_model}, $\cos\theta_\nu = 1$ corresponds to the downward-going neutrino and -1 to the upward-going neutrino. In both these figures, we study six different profiles of the Earth which are i) vacuum, ii) PREM, iii) core-mantle-crust, iv) mantle-crust, v) core-mantle, and vi) uniform density.  

\begin{itemize}
	\item \textbf{Vacuum:} The left panel of the first row in Fig.~\ref{fig:Puu_oscillogram_model} shows the survival probability $P(\nu_{\mu} \rightarrow \nu_{\mu})$ in vacuum where we can identify the first oscillation minimum as a dark blue diagonal band which starts from ($E_\nu = 1$ GeV, $\cos\theta_\nu = 0$) and ends at ($E_\nu = 25$ GeV, $\cos\theta_\nu = -1$). This diagonal band is named as ``oscillation valley''~\cite{Kumar:2020wgz,Kumar:2021lrn}. The higher-order oscillation minima and maxima in vacuum are shown with thinner bands of blue and yellow colors, respectively, in the lower-left triangle. The left panel in the first row in Fig.~\ref{fig:Peu_oscillogram_model} shows $P(\nu_e \rightarrow \nu_\mu)$ for the case of vacuum oscillation where we do not see any matter effect. 	
	
	\item \textbf{PREM profile:} The right panel in the first row in Fig.~\ref{fig:Puu_oscillogram_model} shows the survival probability $P(\nu_\mu \rightarrow \nu_\mu)$ in the presence of matter with PREM profile. The oscillation valley can be observed along with matter effect. The red patch around $-0.8 < \cos\theta_{\nu} < -0.5$ and $6  \text{ GeV} < E_{\nu} < 10  \text{ GeV}$ corresponds to MSW resonance whereas yellow patches around $\cos\theta_{\nu} < -0.8$ and $3 \text{ GeV} < E_{\nu} < 6 \text{ GeV}$ is due to the NOLR/parametric resonance. The right panel in the first row in Fig.~\ref{fig:Peu_oscillogram_model} shows $P(\nu_e \rightarrow \nu_\mu)$ in the presence of matter with PREM profile where we can identify the MSW resonance as single yellow patch around $-0.8 < \cos\theta_\nu < -0.5$ whereas the NOLR/parametric resonance can be seen as two yellow patches around $ \cos\theta_{\nu} < -0.8$. A sharp transition is observed around the boundary of core and mantle at $\cos\theta_\nu = -0.84$.
		
	\item \textbf{Core-mantle-crust profile:} The survival probability $P(\nu_\mu \rightarrow \nu_\mu)$ with the three-layered profile of core-mantle-crust is shown in the left panel of the second row in Fig.~\ref{fig:Puu_oscillogram_model} where we can identify the MSW resonance as well as the NOLR/parametric resonance  similar to the case of PREM profile. The left panel in the second row in Fig.~\ref{fig:Peu_oscillogram_model} shows $P(\nu_e \rightarrow \nu_\mu)$ for core-mantle-crust profile where we can identify the MSW resonance as well as the NOLR/parametric resonance.  Here, we can observe two sharp transitions at core-mantle boundary ($\cos\theta_\nu = -0.84$) and mantle-crust boundary ($\cos\theta_\nu = -0.45$). 
		
	\item \textbf{Mantle-crust profile:} The right panel of the second row in Fig~\ref{fig:Puu_oscillogram_model} shows the survival probability $P(\nu_\mu \rightarrow \nu_\mu)$ for the case of the two-layered profile of mantle-crust where we can observe that the MSW resonance is modified significantly and the NOLR/parametric resonance  is not visible. This indicates that the absence of core modifies the matter effect significantly. Due to the absence of core, the NOLR/parametric resonance  as well as the sharp transition around $\cos\theta_\nu = -0.84$ is absent in $P(\nu_e \rightarrow \nu_\mu)$ for mantle-crust profile as shown in the right panel of the second row in Fig.~\ref{fig:Peu_oscillogram_model}.
		
	\item \textbf{Core-mantle profile:} For the case of the two-layered profile of core-mantle shown in the left panel of the third row in Fig.~\ref{fig:Puu_oscillogram_model}, the MSW resonance, as well as the NOLR/parametric resonance, are observed clearly for the survival probability $P(\nu_\mu \rightarrow \nu_\mu)$ which indicate that the absence of crust does not affect the matter effect by a large amount. For the core-mantle profile shown in the left panel of the third row in Fig.~\ref{fig:Peu_oscillogram_model}, the matter effects for $P(\nu_e \rightarrow \nu_\mu)$  are the same as observed in the case of the three-layered profile, but the sharp transition around $\cos\theta_\nu = -0.45$ is absent because we do not have the mantle-crust boundary in this profile.
	
	\item \textbf{Uniform density:} The right panel of the third row in Fig.~\ref{fig:Puu_oscillogram_model} shows the survival probability $P(\nu_\mu \rightarrow \nu_\mu)$ for the case of uniform density inside Earth where we can identify the MSW resonance, which is disturbed by a small amount. The NOLR/parametric resonance  is absent, which is a sign of the absence of core. In the right panel of the third row of  Fig.~\ref{fig:Peu_oscillogram_model}, $P(\nu_e \rightarrow \nu_\mu)$ is shown for uniform density inside Earth where we can find that the NOLR/parametric resonance, as well as two sharp transitions, are absent. This is because we do not have the core and any boundaries between layers.
\end{itemize}

Thus, we may infer from these plots that the presence of mantle and core results in the MSW resonance and the NOLR/parametric resonance, respectively, whereas boundaries between layers result in sharp transitions. We would like to mention that we have used NO for these plots where a significant matter effect is observed in the neutrino channel, and antineutrinos feel negligible matter effect. If we consider the case of IO, antineutrinos will feel the significant matter effect rather than neutrinos.  Our aim is to observe these features in the reconstructed muon observables at ICAL in 10 years. In Section~\ref{sec:events}, we discuss the method to simulate neutrino events at the ICAL detector. 

\section{Event generation at ICAL}
\label{sec:events}

The 50 kt magnetized ICAL detector at INO~\cite{Kumar:2017sdq} would consist of a stack of iron layers having a thickness of 5.6 cm as a passive detector element with a Resistive Plate Chamber (RPC) sandwiched between them as an active detector element. The charged-current (CC) interactions of neutrinos with iron nuclei result in the production of charged muons. The resulting muon deposits energy in the RPC with the production of signals in the perpendicular strips in X and Y directions that provide (x,\;y) coordinate of hit, whereas the layer number of RPC gives the Z coordinate. Since the multi-GeV muon is a minimum ionizing particle, it passes through many layers and leaves hits in those layers in the form of a track. The charge of muon can be identified by the direction of bending of track in the magnetic field, which results in the ability of ICAL to distinguish between atmospheric neutrinos and antineutrinos in the multi-GeV range of energy. The neutrino interaction is also contributed by resonance scattering and  deep inelastic scattering (DIS) at multi-GeV energy resulting in the production of hadrons. 

In this work, the neutrino interactions are simulated using Monte Carlo (MC) neutrino event generator NUANCE~\cite{Casper:2002sd} using the geometry of ICAL as target and neutrino flux at the proposed INO site~\cite{Athar:2012it,PhysRevD.92.023004} at Theni district of Tamil Nadu, India. The effect of solar modulation on neutrino flux is taken into account by considering flux with high solar activity (solar maximum) for half exposure and low solar activity (solar minimum) for another half. To minimize the statistical fluctuations, we generate 1000-year MC unoscillated neutrino events at ICAL. The three-flavor neutrino oscillations in the presence of matter are taken into account using a reweighting algorithm~\cite{Ghosh:2012px,Thakore:2013xqa,Devi:2014yaa}. 

To incorporate the detector response for muons and hadrons in the current analysis, we have used the look-up tables/migration matrices provided by the ICAL collaboration after performing a rigorous detector simulation study using the widely used GEANT4 package~\cite{Geant4:2003}. The details of these simulation studies performed by the ICAL collaboration are given in Refs.~\cite{Chatterjee:2014vta,Devi:2013wxa}. The Ref.~\cite{Chatterjee:2014vta} discusses in detail how various response functions for muons have been obtained by performing a rigorous GEANT4-based simulation study by the ICAL collaboration. To simulate the detector response, a huge number of muons are passed through the ICAL detector. The muon leaves the hits in various RPC layers in the form of a track. The reconstruction algorithm fits the track using the Kalman filter technique and calculates the vertex, direction, energy, and charge of the muon. The ICAL reconstruction algorithm requires about a minimum of 8 to 10 hits to reconstruct the muon track, which translates to the energy threshold\footnote{The ICAL detector consists of a stack of iron layers with a thickness of 5.6 cm each having a gap of 4 cm between two successive iron layers to insert active RPCs. In the multi-GeV energy range, muon is a minimum ionizing particle and it deposits energy inside a medium at the rate of about $(1/\rho)\cdot (dE/dx) \sim 2 \text{ MeV\,g}^{-1}\text{\,cm}^2$ as described in the PDG~\cite{Zyla:2020zbs}. For the case of iron ($\rho \sim 7.9$ g/cm\textsuperscript{3}), the muon in the GeV energy range will deposit energy of about 16 MeV/cm, and this will lead to an energy loss of about 100 MeV in each layer of iron (thickness of 5.6 cm) in ICAL. We need about a minimum of 8 to 10 hits to reconstruct a muon track at ICAL, which results in an energy threshold of about 1 GeV.} of about 1 GeV. The outcomes of migration matrices are nicely summarized as a function of input muon momentum for various input zenith angles in Ref.~\cite{Chatterjee:2014vta}. The authors in Ref.~\cite{Chatterjee:2014vta} show reconstruction efficiency in Fig.~13, charge identification (CID) efficiency in Fig.~14, muon energy resolution in Fig.~11, and muon angular resolution in Fig.~6. 

The reconstruction efficiency increases sharply with the input muon energy up to 2 GeV, and then it saturates to around  80\% to 90\% in the muon energy range of 2 to 20 GeV for a wide range of zenith angle starting from $\cos\theta_\mu = 0.35$ to 0.85 as shown in Fig.~13 of  Ref.~\cite{Chatterjee:2014vta}. In ICAL, the number of events is less in the horizontal direction because our reconstruction efficiency is poor in this case due to the horizontally stacked layers of Resistive Plate Chamber (RPC) where only a few RPC layers receive hits in case of horizontal events. As far as the charge identification is concerned, the ICAL detector is expected to perform quite well in the muon energy range of 1 to 20 GeV since it plans to have a magnetic field of around 1.5 T which will be sufficient enough to get the curvature of the muon track to identify the charge of the muon. The charge identification efficiency at ICAL is about 98\% in the muon energy range of 2 to 20 GeV for various zenith angles in the range of $\cos\theta_\mu = 0.35$ to 0.85 as shown in Fig.~14 of Ref.~\cite{Chatterjee:2014vta}. The reconstructed muon energies and directions are fitted with Gaussian distribution to calculate means and standard deviations. The standard deviation represents the detector resolution of the reconstructed parameter. Figure~11 in Ref.~\cite{Chatterjee:2014vta} portrays that the muon energy resolution of the ICAL detector in the muon energy range of 2 to 20 GeV for zenith angles in the range of $\cos\theta_\mu = 0.35$ to 0.85 is approximately 10 to 15\%, which is sufficient enough to capture the information about neutrino oscillation parameters and Earth's matter effect in the multi-GeV energy range for a wide range of baselines. The ICAL detector has an excellent angular resolution of less than $1^\circ$ for a large muon energy range and a wide range of zenith angles, as shown in Fig.~6 of Ref.~\cite{Chatterjee:2014vta}. These numbers tell us that the ICAL detector performs quite well as far as the reconstruction of the four-momenta of muon is concerned, which is important to have the sensitivity of ICAL towards the structure of Earth.

Now, let us elaborate on how the ICAL collaboration obtains the hadron energy response inside the ICAL detector. In the multi-GeV range of energies, the resonance scattering and deep inelastic scattering of neutrinos produce hadrons along with muons. Unlike muons, hadrons produce multiple hits in a single layer of RPC, and this leads to shower-like events. These hadrons take away a significant fraction of the incoming neutrino energy, and the hadron energy deposited in the detector is defined using a variable ${E'}_\text{had} = E_\nu - E_\mu$ in Ref~\cite{Devi:2013wxa}. This Reference discusses in detail how the hadron energy resolution has been obtained by performing a rigorous GEANT4-based simulation study by the ICAL collaboration. The hadron energy response is simulated by passing a huge number of hadrons through ICAL geometry. The distribution of the total number of hits for these hadrons is fitted with the Vavilov distribution function. The mean number of hits and the square root of the variance obtained after fitting is related to the energy and the energy resolution of hadron, respectively. Figure~8 in Ref.~\cite{Devi:2013wxa} shows that the hadron energy resolution is about 40 to 60\% in the energy range of 2 to 8 GeV and about 40\% for energies above 8 GeV. Though the hadron energy resolution is not as good as muon, it is sufficient to capture the possible correlation between four-momenta of muon ($E_\mu^\text{rec}$, $\cos\theta_\mu^\text{rec}$) and hadron energy (${E'}_\text{had}^\text{rec}$) which we treat as independent variables.

\begin{table}[htb!]
	\begin{center}
		\begin{tabular}{|l|c|c|c|c|c|c|}
			\hline \hline
			\multirow{2}{*}{Profiles} & \multicolumn{3}{c|}{Reconstructed $\mu^-$ events}  & \multicolumn{3}{c|}{Reconstructed $\mu^+$ events}\\ \cline{2-7}
			& Upward & Downward & Total & Upward & Downward & Total \\
			\hline
			PREM & 1654 & 2960 & 4614 & 741 & 1313 & 2053 \\
			Core-Mantle-Crust & 1659 & 2960 & 4619 & 739 & 1313 & 2052 \\
			Vacuum & 1692 & 2960 & 4652 & 745 & 1313 & 2057 \\
			\hline \hline
		\end{tabular}
		\mycaption{The total number of reconstructed $\mu^-$ and $\mu^+$ events expected in the upward and downward direction at the 50 kt ICAL detector in 10 years which is scaled from 1000-year MC data. We take the three-flavor oscillation parameters from Table~\ref{tab:osc-param-value}. We assume NO and $\sin^2\theta_{23} = 0.5$.
		}
		\label{tab:events}
	\end{center}
\end{table}

After folding with these detector properties following the procedure mentioned in~\cite{Ghosh:2012px,Thakore:2013xqa,Devi:2014yaa}, we obtain reconstructed $\mu^-$ and $\mu^+$ events at ICAL. These reconstructed events for 1000-year MC are then scaled to 10-year MC. For the case of NO, the 50 kt ICAL detector would detect about 4614 reconstructed $\mu^-$ and 2053 reconstructed $\mu^+$ events in 10 years with a total exposure of 500 kt$\cdot$yr using three-flavor neutrinos oscillation with matter effect considering 25-layered PREM profile of Earth as shown in Table~\ref{tab:events}. The ns timing resolution of RPCs~\cite{Dash:2014ifa,Bhatt:2016rek,Gaur:2017uaf} enables ICAL to distinguish between upward-going and downward-going muon events. ICAL is expected to detect about 1654 upward-going and 2960 downward-going $\mu^-$ events in 10 years, whereas $\mu^+$ events in upward and downward direction will be around 741 and 1313, respectively, in 10 years. Table~\ref{tab:events} also shows events considering the three-layered profile of core-mantle-crust as well as the vacuum where we can observe some difference in upward-going events only which have experienced matter effect. It is important to note that there is not much difference in total event rate for these profiles, but the final result receives a contribution from binning of these events, which is possible because of good resolution of energy and direction of reconstructed muons at ICAL. For the case of NO, about 98\% of $\mu^-$ and 99\% of $\mu^+$ events at ICAL are contributed by survival channel $(\nu_\mu \rightarrow \nu_\mu)$ whereas the remaining contribution is from appearance channel $(\nu_e \rightarrow \nu_\mu)$.

The direction of these reconstructed muons can be used to get information about the regions in the Earth through which neutrino has traversed as discussed in Section~\ref{sec:events_layer}.

\section{Identifying events passing through different layers of Earth}
\label{sec:events_layer}

\begin{figure}[htb!]
	\centering
	\includegraphics[width=0.6\linewidth]{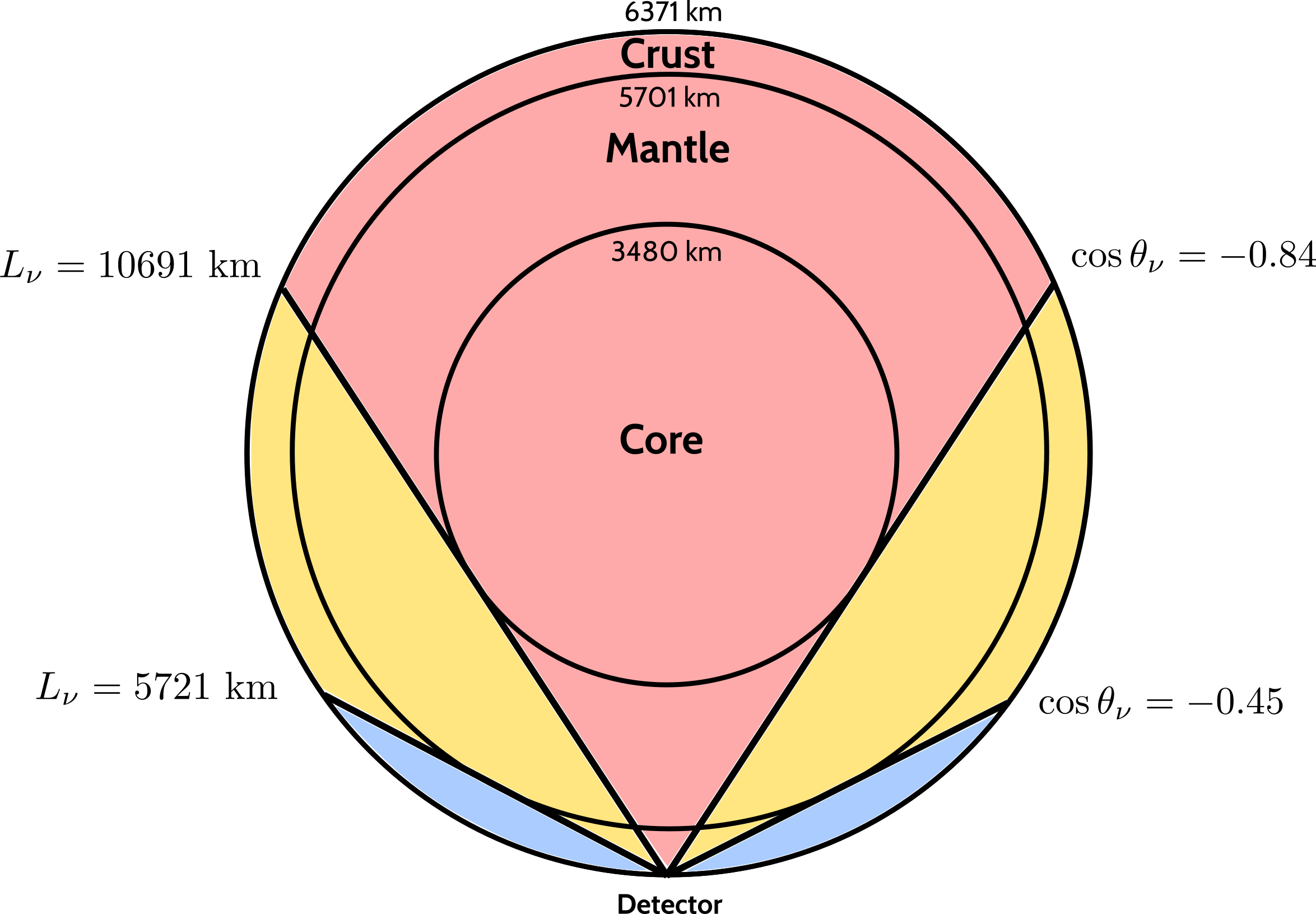}
	\mycaption{Neutrinos passing through regions consisting of a particular set of layers of Earth depending upon the zenith angle.
	}
	\label{fig:neutrino-path}
\end{figure}

The atmospheric neutrinos cover a wide range of baselines\footnote{The neutrino baseline $L_\nu$ is related to the neutrino zenith angle $\theta_\nu$ by 
\begin{equation}
L_\nu = \sqrt{(R+h)^2 - (R-d)^2\sin^2\theta_\nu} \,-\, (R-d)\cos\theta_\nu \, ,
\end{equation}
where, $R$, $h$, and $d$ correspond to the radius of Earth, the average height from the surface of Earth at which neutrinos are produced, and the depth of the detector below the surface of Earth, respectively. In our analysis, we use $R = 6371$ km, $h = 15$ km, and $d = 0$ km. 
} ($L_\nu$) 
from 15 km to 12757 km that correspond to downward and upward directions, respectively. The vertically upward-going neutrinos pass through a set of layers of Earth depending upon their direction as shown in Fig.~\ref{fig:neutrino-path}. The vertically upward-going neutrinos with $\cos\theta_\nu < -0.84$ pass through crust-mantle-core region as shown by pink color in Fig.~\ref{fig:neutrino-path}. The yellow region in Fig.~\ref{fig:neutrino-path} with $-0.84 <\cos\theta_\nu < -0.45$ shows the neutrino events passing through crust-mantle region. The neutrinos which pass through only crust are shown by the blue color region in Fig.~\ref{fig:neutrino-path}.

\begin{table}[htb!]
	\centering
	\begin{tabular}{|c|c|c|c|c|}
		\hline\hline
		Regions & $\cos\theta_{\nu}$ & $L_\nu$ (km) & $\mu^{-}$ Events & $\mu^{+}$ Events\\
		\hline
		Crust-mantle-core&	(-1.00, -0.84) & (10691, 12757) & 331 & 146 \\
		Crust-mantle & (-0.84, -0.45) & (5721, 10691) & 739 & 339\\
		Crust & (-0.45, 0.00) & (437, 5721) & 550 & 244 \\ 
		Downward & (0.00, 1.00) & (15, 437) & 2994 & 1324 \\ 
		Total & (-1.00, 1.00) & (15, 12757) & 4614 & 2053 \\ 
		\hline\hline
	\end{tabular}
	\mycaption{Reconstructed $\mu^-$ and $\mu^+$ events expected at ICAL for 500 kt$\cdot$yr exposure for neutrinos passing through various regions depending upon zenith angle of neutrino. These reconstructed muon events for 10 years are scaled from 1000-year MC data. We consider three-flavor neutrino oscillations in the presence of matter with the PREM profile. We take the three-flavor oscillation parameters from Table~\ref{tab:osc-param-value}. We assume NO and $\sin^2\theta_{23} = 0.5$.}
	\label{tab:layer-passing-event}
\end{table}

Table~\ref{tab:layer-passing-event} shows the expected number of events at ICAL for 500 kt$\cdot$yr exposure for neutrinos passing through different regions shown in Fig.~\ref{fig:neutrino-path}. Here, we consider three-flavor neutrino oscillations in the presence of matter with the PREM profile of Earth. ICAL would detect about 331 (146) $\mu^-$ ($\mu^+$) events corresponding to the crust-mantle-core passing neutrinos (antineutrino). About 739 $\mu^-$ and 339 $\mu^+$ events would be detected for crust-mantle passing neutrinos and antineutrinos, respectively. The events passing through only crust would be about 550 and 244 for $\mu^-$ and $\mu^+$, respectively. 

Note that the total number of events for reconstructed $\mu^-$ (4614) and $\mu^+$ (2053) are the same in Table~\ref{tab:layer-passing-event} and Table~\ref{tab:events} for the PREM profile case. Also, it is worthwhile to mention that the reconstructed downward-going $\mu^-$ and $\mu^+$ events mentioned in Table~\ref{tab:layer-passing-event} are a bit different as compared to the downward-going events as mentioned in Table~\ref{tab:events} for the PREM profile case. It happens because of the angular smearing caused by kinematics and finite angular resolution of the detector. Because of this angular smearing, we may have differences in the direction of neutrinos and reconstructed muons, which may force a downward-going neutrino (near horizon) to appear as an upward-going reconstructed muon event.

\begin{figure}[htb!]
	\centering
	\includegraphics[width=0.44\linewidth]{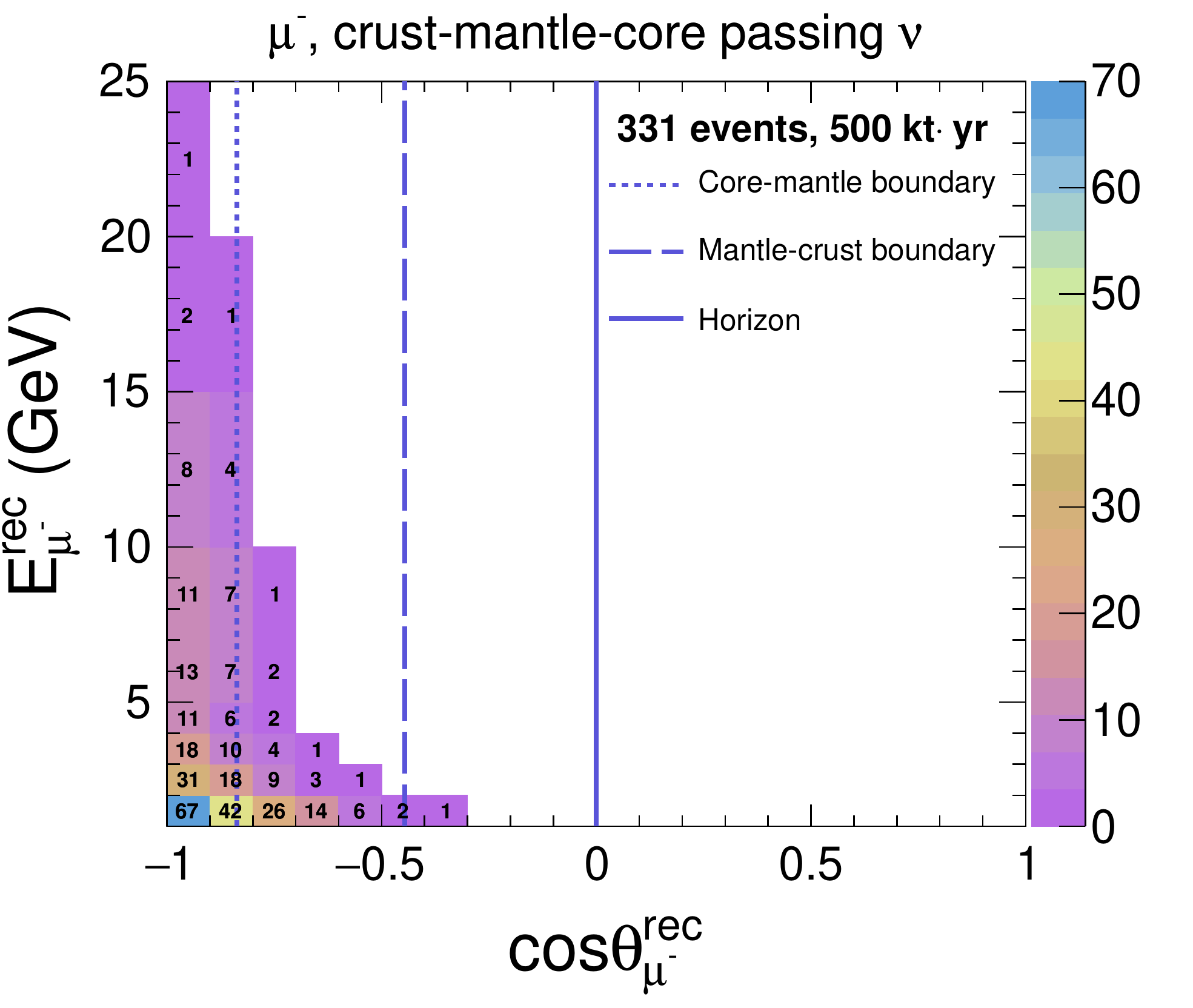}
	\includegraphics[width=0.44\linewidth]{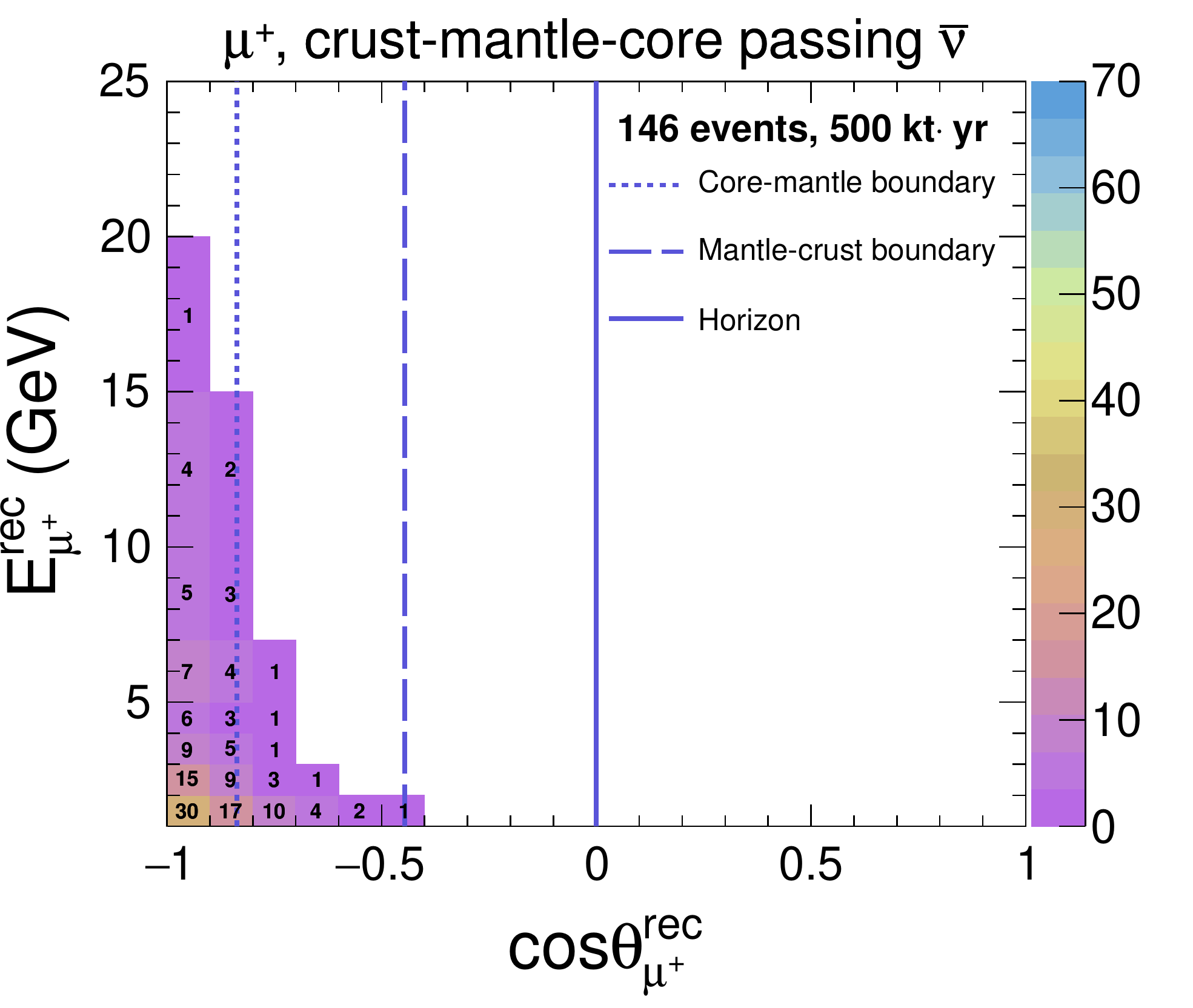}
	\includegraphics[width=0.44\linewidth]{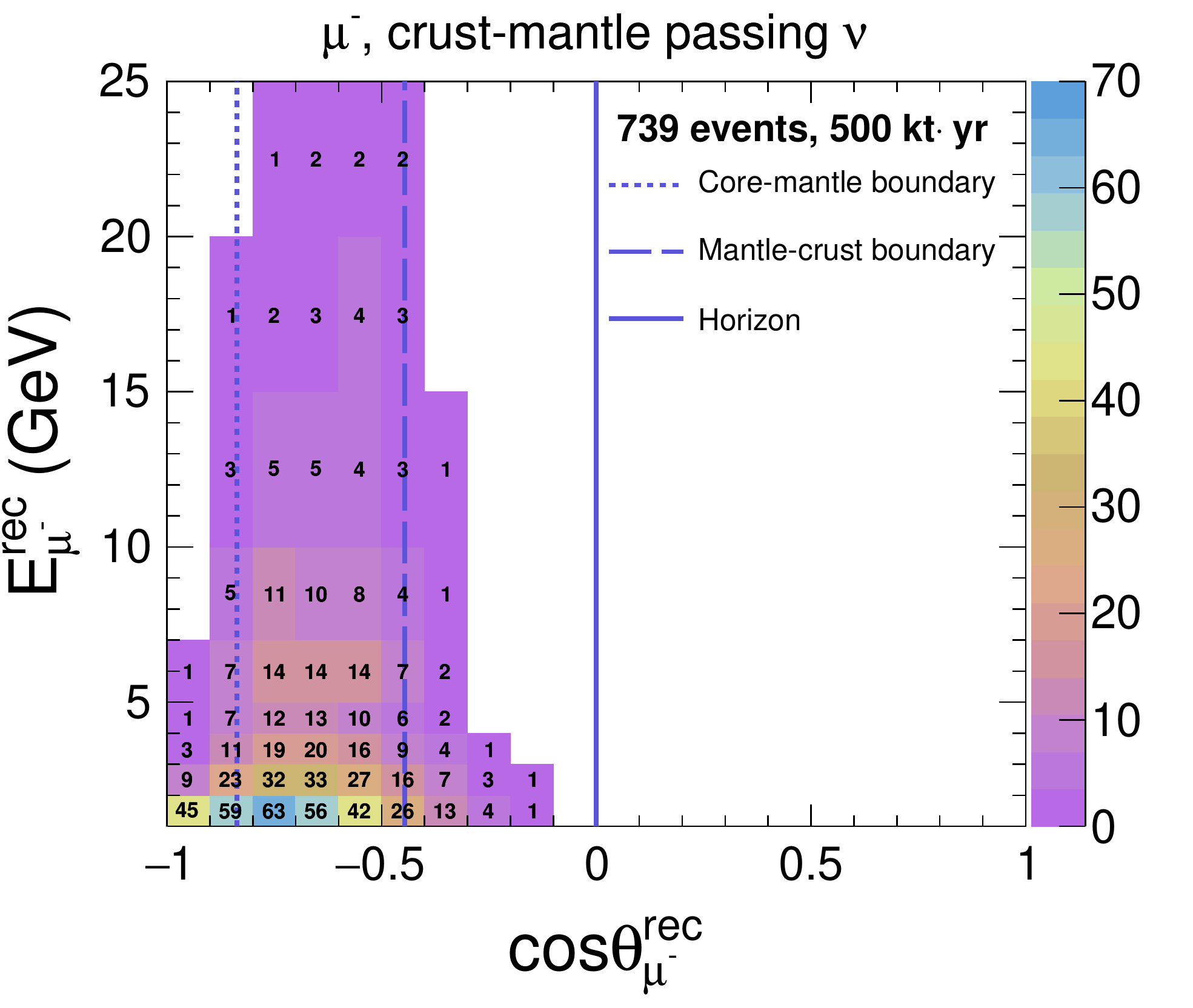}
	\includegraphics[width=0.44\linewidth]{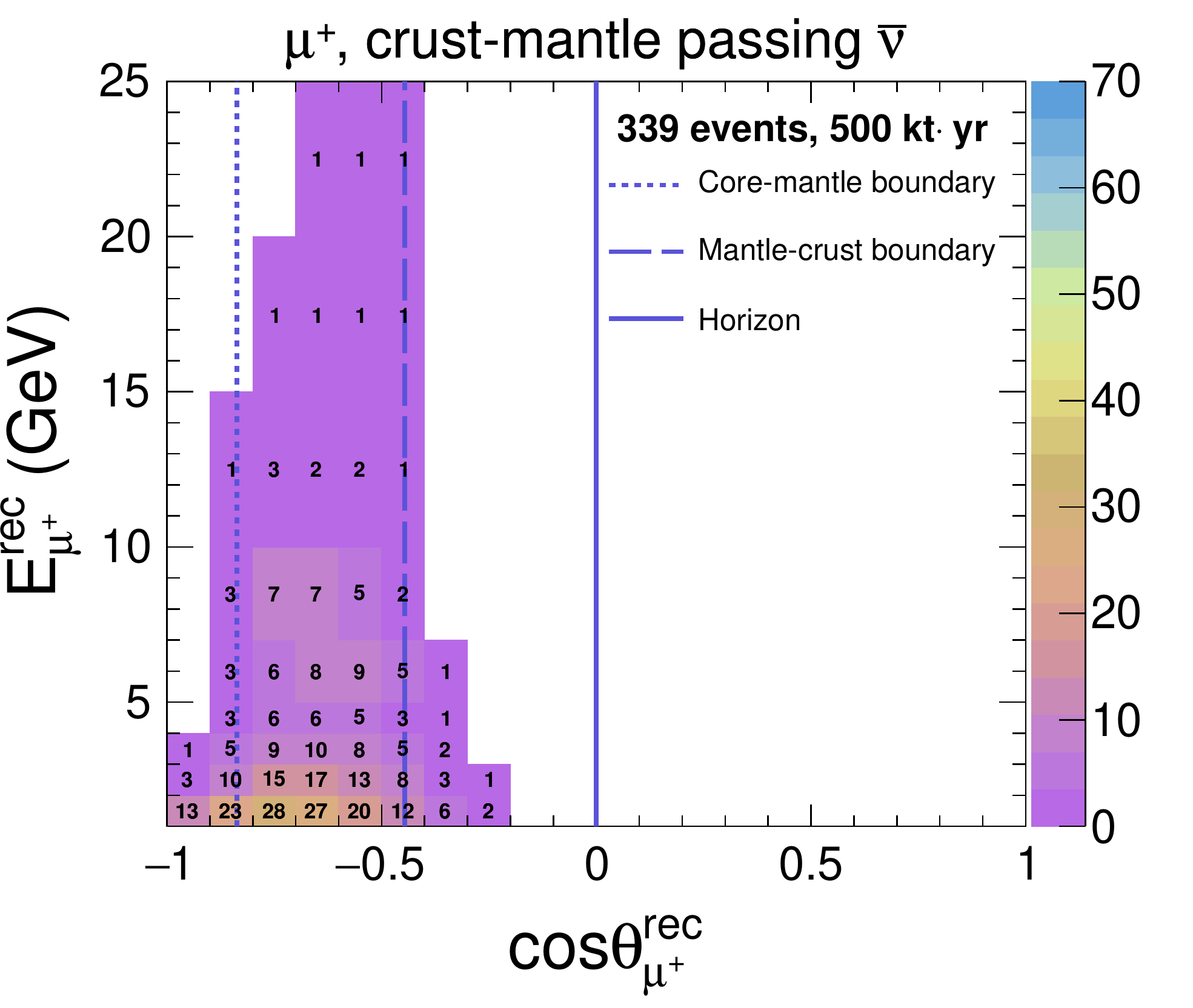}
	\includegraphics[width=0.44\linewidth]{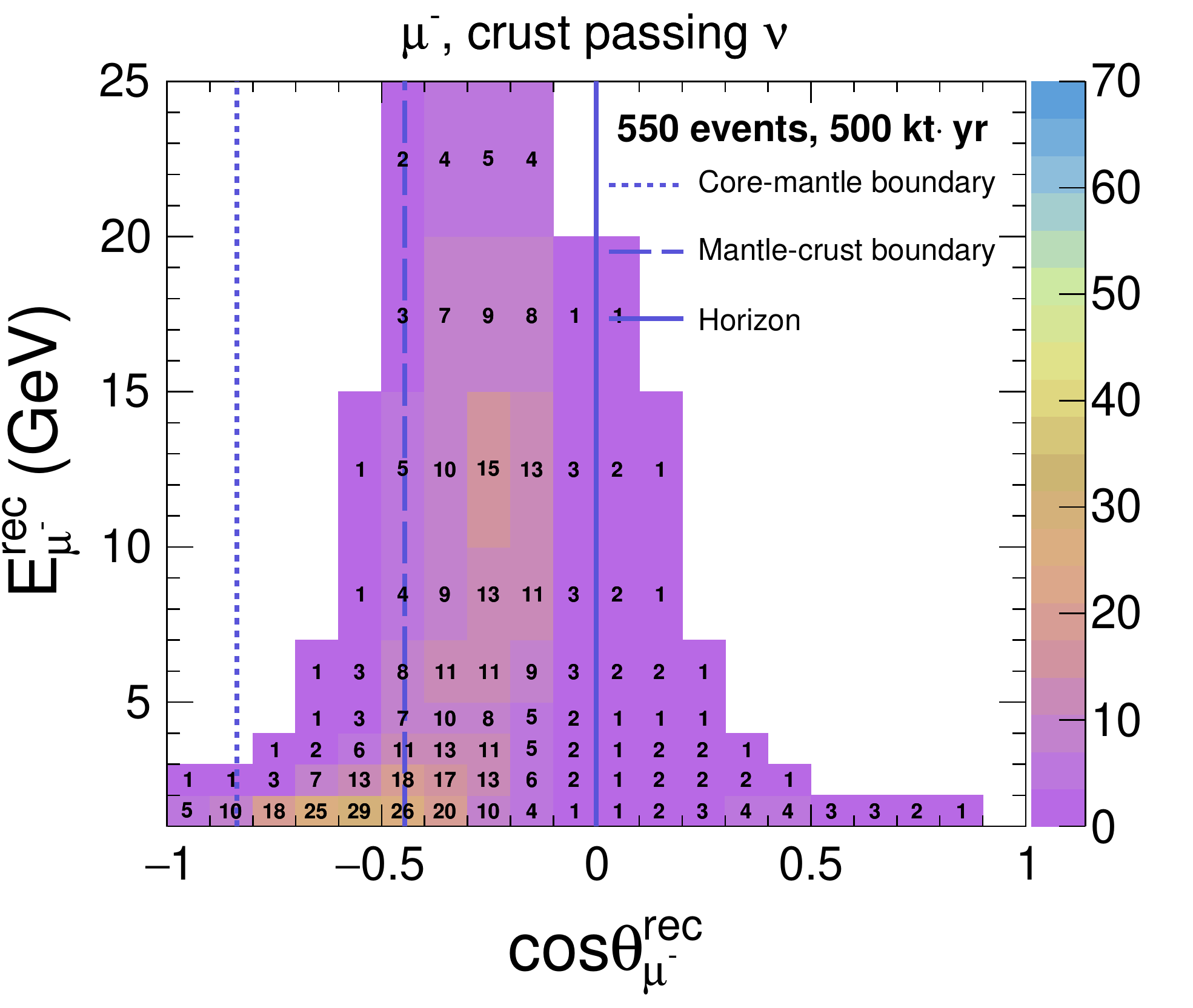}
	\includegraphics[width=0.44\linewidth]{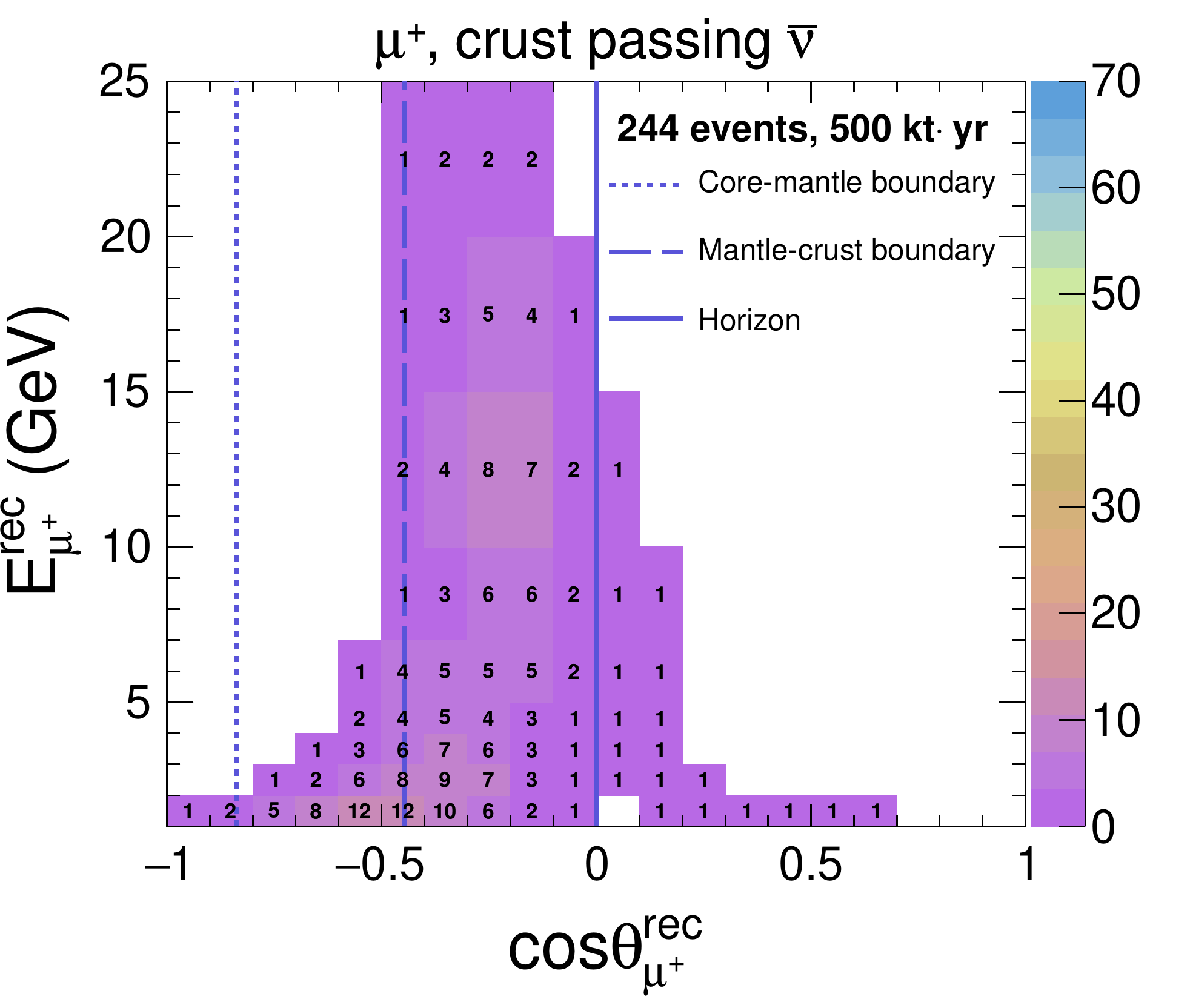}
	\mycaption{Reconstructed muon event distribution at ICAL for 500 kt$\cdot$yr exposure for neutrinos traversed through various regions in the Earth. These reconstructed muon events for 10 years are scaled from 1000-year MC data. We consider three-flavor neutrino oscillations in the presence of matter with the PREM profile. We take the three-flavor oscillation parameters from Table~\ref{tab:osc-param-value}. We assume NO and $\sin^2\theta_{23} = 0.5$. The top, middle, and bottom panels show the distribution of reconstructed muon events for the parent neutrinos passing through the crust-mantle-core, crust-mantle, and crust, respectively. The dotted, dashed, and solid vertical blue lines correspond to the core-mantle boundary, mantle-crust boundary, and horizontal direction, respectively. The Left (right) panel shows reconstructed $\mu^-$ ($\mu^+$) events.}
	\label{fig:event_dist_nu_density_zones}
\end{figure}

We would like to point out that while using reconstructed muon observables, the difference in the direction of muon and neutrino due to angular smearing may cause a deterioration in the capability of ICAL to identify the region traversed by neutrino. Figure~\ref{fig:event_dist_nu_density_zones} shows event distribution of reconstructed muons in ($E_\mu^\text{rec}$, $\cos\theta_\mu^\text{rec}$) plane for neutrinos passing through different regions. For demonstrating reconstructed event distribution for 500 kt$\cdot$yr exposure, we have chosen a binning scheme such that we have total 9 bins in $E_\mu^\text{rec}$ and 20 bins in $\cos\theta_\mu^\text{rec}$. For $E_\mu^\text{rec}$, we have 5 bins of 1 GeV in the range 1 -- 5 GeV, 1 bin of 2 GeV in the range 5 -- 7 GeV, 1 bin of 3 GeV in the range 7 -- 10 GeV, and 3 bins of 5 GeV in the range 10 -- 25 GeV, whereas uniform bins of 0.1 is used for $\cos\theta_\mu^\text{rec}$ in the range of -1 to 1. In Fig.~\ref{fig:event_dist_nu_density_zones}, the vertical dotted blue line shows the core-mantle boundary whereas vertical dashed blue line shows the mantle-crust boundary. The horizontal direction is shown with solid blue line.

The left panel in the first row in Fig.~\ref{fig:event_dist_nu_density_zones} shows distribution of reconstructed $\mu^-$ events for neutrinos passing through crust-mantle-core region. Here, we can observe that although the actual neutrinos are present only on the left side of the dotted blue line, few reconstructed muons get smeared into other regions also. The left panel in the second row in Fig.~\ref{fig:event_dist_nu_density_zones} shows event distribution of reconstructed $\mu^-$ events for neutrinos passing through crust-mantle region \textit{i.e.} between dotted and dashed blue lines. Although most of the events remain between dotted and dashed blue lines, some events smear into other regions also. The reconstructed $\mu^-$ events distribution for crust passing neutrinos is shown in the left panel of the third row in Fig.~\ref{fig:event_dist_nu_density_zones}. A similar kind of smearing is observed for the distribution of reconstructed $\mu^+$ events as shown in the right panels in Fig.~\ref{fig:event_dist_nu_density_zones}. Thus, we can say that the good directional resolution at ICAL enables the reconstructed muon events to preserve the information about regions traversed by neutrinos.

\section{Simulation method}
\label{sec:statistical analysis}

\subsection{Binning scheme}

\begin{table}[htb!]
	\centering
	\begin{tabular}{|c|c|c|c c|}
		\hline \hline
		Observable & Range & Bin width & \multicolumn{2}{c|}{Number of bins} \\
		\hline
		\multirow{4}{*}{ $E_\mu^{\rm rec}$ (GeV)} & [1, 4] &  0.5 & 6 & \rdelim\}{4}{7mm}[12] \cr 
		& [4, 7] & 1 & 3  & \cr
		& [7, 11] & 4 & 1  & \cr
		& [11, 21] & 5 & 2  & \cr
		\hline 
		\multirow{3}{*}{$\cos\theta_\mu^{\rm rec}$} & [-1.0, -0.4] & 0.05 & 12  & \rdelim\}{3}{7mm}[21]\cr
		& [-0.4, 0.0] & 0.1 & 4  & \cr
		& [0.0, 1.0] & 0.2 & 5  & \cr
		\hline 
		\multirow{3}{*}{ ${E'}_\text{had}^{\rm rec}$ (GeV)} & [0, 2] &  1 & 2 & \rdelim\}{3}{7mm}[4] \cr 
		& [2, 4] & 2 & 1  & \cr
		& [4, 25] & 21 & 1  & \cr
		\hline \hline
	\end{tabular}
	\mycaption{The binning scheme considered for reconstructed observables $E_\mu^{\rm rec}$, $\cos\theta_\mu^{\rm rec}$, and ${E'}_\text{had}^{\rm rec}$ for $\mu^-$ as well as $\mu^+$ events. }
	\label{tab:binning-2D-10years}
\end{table}

In this work, we are harnessing the matter effect to understand the distribution of matter inside the Earth. The binning scheme used in Ref.~\cite{Devi:2014yaa} is optimized to probe the Earth's matter effect considering $E_\mu^\text{rec}$, $\cos\theta_\mu^\text{rec}$, and ${E'}_\text{had}^\text{rec}$ as observables. In this binning scheme, $E_\mu^\text{rec}$ is considered in the range of 1 -- 11 GeV whereas ${E'}_\text{had}^\text{rec}$ is having a range of 0 -- 15 GeV. We have modified this binning scheme by adding two bins of 5 GeV for $E_\mu^\text{rec}$ in the range  11 -- 21 GeV whereas last bin of ${E'}_\text{had}^\text{rec}$ is increased up to 25 GeV. The resulting binning scheme is shown in Table~\ref{tab:binning-2D-10years} where we have total 12 bins in $E_\mu^\text{rec}$, 21 bins in $\cos\theta_\mu^\text{rec}$ and 4 bins in ${E'}_\text{had}^\text{rec}$. We would like to mention that the bin sizes are chosen following the detector resolutions such that there is a sufficient number of events in each bin. Although, matter effect is experienced by upward-going neutrinos only, we have considered $\cos\theta_\mu^\text{rec}$ in the range of -1 to 1 because downward-going events help in increasing overall statistics as well as minimizing normalization uncertainties in atmospheric neutrino events. This also incorporates those upward-going (near horizon) neutrino events that result in downward-going reconstructed muon events due to angular smearing during neutrino interaction as well as reconstruction. We have considered the same binning scheme for $\mu^-$ as well as $\mu^+$.

\subsection{Numerical analysis}

In this analysis, the $\chi^2$ statistics is expected to give median sensitivity of the experiment in the frequentist approach~\cite{Blennow:2013oma}. We define the following Poissonian $\chi^2_{-}$ for $\mu^-$ in $E_\mu^\text{rec}$, $\cos\theta_\mu^\text{rec}$, and ${E'}_\text{had}^\text{rec}$ observables as considered in Ref.~\cite{Devi:2014yaa}:

\begin{equation}\label{eq:chisq_mu-}
\chi^2_- = \mathop{\text{min}}_{\xi_l} \sum_{i=1}^{N_{{E'}_\text{had}^\text{rec}}} \sum_{j=1}^{N_{E_{\mu}^\text{rec}}} \sum_{k=1}^{N_{\cos\theta_\mu^\text{rec}}} \left[2(N_{ijk}^\text{theory} - N_{ijk}^\text{data}) -2 N_{ijk}^\text{data} \ln\left(\frac{N_{ijk}^\text{theory} }{N_{ijk}^\text{data}}\right)\right] + \sum_{l = 1}^5 \xi_l^2
\end{equation}

where, 
\begin{equation}
N_{ijk}^\text{theory} = N_{ijk}^0\left(1 + \sum_{l=1}^5 \pi^l_{ijk}\xi_l\right)
\end{equation}

$N_{ijk}^\text{theory}$ and $N_{ijk}^\text{data}$ represent the expected and observed number of events for $\mu^-$ in a given $(E_\mu^\text{rec}, \cos\theta_\mu^\text{rec}, {E'}_\text{had}^\text{rec})$ bin whereas $N_{ijk}^0$ are the number of events without considering systematic uncertainties . In this analysis, we use the method of pulls~\cite{GonzalezGarcia:2004wg,Huber:2002mx,Fogli:2002} to consider five systematic uncertainties following Refs.~\cite{Ghosh:2012px,Thakore:2013xqa}: flux normalization error (20\%), cross section error (10\%), energy dependent tilt error in flux (5\%), error in zenith angle dependence of flux (5\%), and overall systematics (5\%).

Following the same procedure, we define $\chi^2_{+}$ for $\mu^+$ which will be calculated separately along with $\chi^2_{-}$. The total $\chi^2_\text{ICAL}$ for ICAL is calculated by adding $\chi^2_{-}$ and $\chi^2_{+}$.
\begin{equation}\label{eq:chisq_total}
\chi^2_\text{ICAL} = \chi^2_{-} + \chi^2_+.
\end{equation}
We use the benchmark choice of oscillation parameters given in Table~\ref{tab:osc-param-value} as true parameters for simulating data. In theory, first of all, the $\chi^2_\text{ICAL}$ is minimized with respect to pull variables $\xi_l$ and then, marginalization is done for oscillation parameters  $\sin^2\theta_{23}$ in the range (0.36, 0.66), $\Delta m^2_\text{eff}$ in the range (2.1, 2.6) $\times10^{-3}$ eV\textsuperscript{2}, and mass ordering over NO and IO. The solar oscillation parameters $\sin^2 2\theta_{12}$ and $\Delta m^2_{21}$ are kept fixed at their true values given in Table~\ref{tab:osc-param-value} while performing the fit. As far as the reactor mixing angle is concerned, we consider a fixed value of $\sin^2 2\theta_{13} = 0.0875$ both in data and theory since this parameter is already very well measured~\cite{Marrone:2021,NuFIT,Esteban:2020cvm,deSalas:2020pgw}.
Throughout this analysis, we consider $\delta_{\rm CP} = 0$ both in data and theory.

\section{Results}
\label{sec:results}

For statistical analysis, we simulate the prospective data assuming the three-layered core-mantle-crust profile as the true profile of the Earth. The statistical significance of the analysis for ruling out the mantle-crust profile with respect to the core-mantle-crust profile is quantified in the following way
\begin{equation}\label{eq:chisq_diff}
\Delta \chi^2_\text{ICAL-profile} = \chi^2_\text{ICAL}~ (\text{mantle-crust}) - \chi^2_\text{ICAL}~ (\text{core-mantle-crust})
\end{equation}
where, $\chi^2_\text{ICAL}$ (mantle-crust) and  $\chi^2_\text{ICAL}$  (core-mantle-crust) is calculated by fitting prospective data with mantle-crust profile and core-mantle-crust profile, respectively. Since the statistical fluctuations are suppressed,  we have $\chi^2_\text{ICAL}~ (\text{core-mantle-crust}) \sim 0$. 

\subsection{Effective regions in $(E_\mu^\text{rec}, \cos\theta_\mu^\text{rec})$ plane to validate Earth's core}
\label{sec:results_core}

The sensitivity of ICAL towards various density profiles of Earth mainly stems from the Earth's matter effect experienced by neutrinos and antineutrinos while they travel long distances inside the Earth. For a given mass ordering, the Earth's matter effects felt by neutrinos and antineutrinos are different, which in turn alter the neutrino and antineutrino oscillation probability in a different fashion. In this work, while distinguishing between various density profiles of the Earth, the major part of the sensitivity comes from neutrino (antineutrino) mode if the true mass ordering is assumed to be NO (IO). We have elaborated on this issue in the next paragraph. On the contrary, while determining the sensitivity of ICAL towards neutrino mass ordering, both neutrino and antineutrino events contribute irrespective of the choice of true mass ordering~\cite{Ghosh:2012px,Devi:2014yaa}. 

The sensitivity of ICAL to rule out the simple two-layered mantle-crust profile of the Earth in theory while generating the prospective data with the three-layered core-mantle-crust profile mostly comes from $\mu^-$ ($\mu^+$) events if the true mass ordering is NO (IO). In the fixed-parameter scenario, we obtain the median $\Delta \chi^2_\text{ICAL-profile}$ (Data: core-mantle-crust, theory: mantle-crust) of 6.90 (4.10) if the true mass ordering is NO (IO). Note that when NO is our true choice, the contribution towards the fixed-parameter $\Delta \chi^2$ from $\mu^-$ ($\mu^+$) events is 6.85 (0.05). We see a completely opposite trend when IO is our true choice. For the true IO scenario, the contribution towards the fixed-parameter $\Delta \chi^2$ from $\mu^-$ ($\mu^+$) events is 0.02 (4.08).

\begin{figure}
	\centering
	\includegraphics[width=0.48\linewidth]{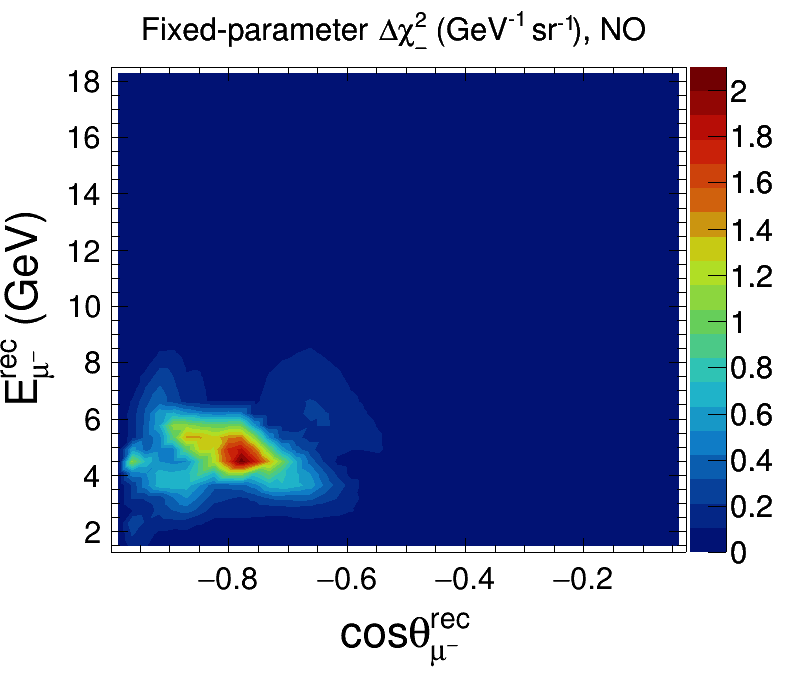}
	\includegraphics[width=0.48\linewidth]{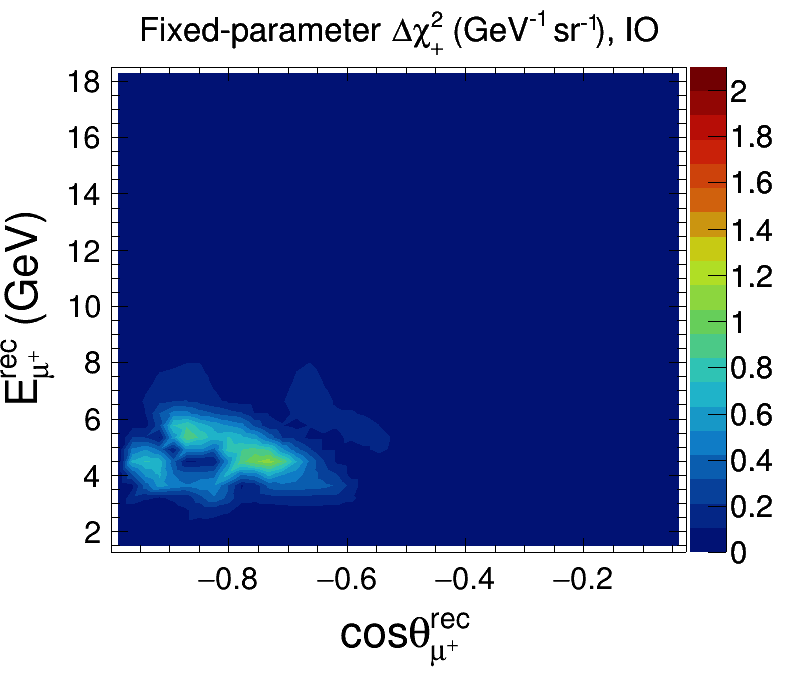}
	\mycaption{The distribution of fixed-parameter $\Delta \chi^2_{-}$ ($\Delta \chi^2_+$) with NO (IO) without pull penalty term for ruling out the mantle-crust profile in theory w.r.t. the core-mantle-crust profile in data in ($E_\mu^\text{rec},\; \cos\theta_\mu^\text{rec}$) plane as shown in the left (right) panel. Note that $\Delta \chi^2_{-}$ and $\Delta \chi^2_{+}$ is plotted in the unit of GeV\textsuperscript{-1} sr\textsuperscript{-1} where we have divided them by $2\pi\times \text{bin area}$.
	The $\Delta \chi^2_-$ ($\Delta \chi^2_+$) for IO (NO) is not significant, and hence, not shown here. We take the three-flavor oscillation parameters from Table~\ref{tab:osc-param-value}.}
	\label{fig:chisq_contour}
\end{figure}

To identify the ranges of energy and direction which are contributing significantly to $\Delta \chi^2_\text{ICAL-profile}$ for ruling out the two-layered mantle-crust profile in theory against the three-layered core-mantle-crust profile in data, we have plotted the distribution of fixed-parameter $\Delta \chi^2_{-}$ and $\Delta \chi^2_+$ without pull penalty term\footnote{$\Delta \chi^2_{-}$ and $\Delta \chi^2_+$ are calculated without pull penalty $\sum_{l = 1}^5 \xi_l^2$ (see Eq.~\ref{eq:chisq_mu-}) to explore contributions from each bin in  ($E_\mu^\text{rec},\; \cos\theta_\mu^\text{rec}$) plane for $\mu^-$ and $\mu^+$ events, respectively.} as the contribution  towards $\Delta \chi^2$ from $\mu^-$ and $\mu^+$ events, respectively in ($E_\mu^\text{rec},\;\cos\theta_\mu^\text{rec}$) plane as shown in Fig.~\ref{fig:chisq_contour}. The left panel of Fig.~\ref{fig:chisq_contour} shows the distribution of $\Delta \chi^2_{-}$ (GeV\textsuperscript{-1} sr\textsuperscript{-1}) for NO in the plane of ($E_\mu^\text{rec},\;\cos\theta_\mu^\text{rec}$) where we can observe that the sensitivity to rule out Earth's core is contributed significantly by bins of higher baselines and multi-GeV energies in the range of 3 to 7 GeV of the reconstructed muons. The baselines with significant contribution correspond to the region around the boundary of core and mantle, where the matter density gets modified significantly during the merger of core and mantle to form the two-layered profile being probed here. We would like to mention that the detector response is already optimized by the ICAL collaboration for these core-passing events in the above-mentioned multi-GeV energy range as described in Section~\ref{sec:events}. Since the reconstructed muon energy threshold of 1 GeV is much lower than the energies contributing to the sensitivity of ICAL toward the Earth's matter effect, the sensitivity of ICAL towards validating Earth's core is not going to be affected by the possible fluctuations around the energy threshold of 1 GeV in the ICAL detector. The contribution of $\Delta \chi^2_{+}$ for NO is negligible and hence not shown here. In the same fashion, the right panel of Fig.~\ref{fig:chisq_contour} shows the distribution of $\Delta \chi^2_{+}$ (GeV\textsuperscript{-1} sr\textsuperscript{-1}) for IO where also, the contribution appears for the lower energy and higher baseline. The contribution of $\Delta \chi^2_{+}$ for IO is smaller than that for $\Delta \chi^2_{-}$ for NO because the lower cross-section for antineutrino results in the lesser statistics of $\mu^+$ events compared to $\mu^-$ events. For the case of IO, the contribution of $\Delta \chi^2_{-}$ is not significant.  

\subsection{Sensitivity to validate Earth's core with and without CID}
\label{sec:results_core_margin}

\begin{table}[hbt!]
	\begin{center}
		\begin{tabular}{|c|c|c|c|c|c|}
			\hline \hline
			\multirow{3}{*}{MC Data} & \multirow{3}{*}{Theory} & \multicolumn{4}{c|}{$\Delta \chi^2_\text{ICAL-profile}$}\\ \cline{3-6}
			& & \multicolumn{2}{c|}{NO(true)} & \multicolumn{2}{c|}{IO(true)}\\ \cline{3-6}
			& & with CID & w/o CID & with CID & w/o CID\\
			\hline
			Core-mantle-crust & Vacuum    & 4.65 & 2.96 & 3.53 & 1.43 \\
			Core-mantle-crust & Mantle-crust& 6.31 & 3.19 & 3.92 & 1.29 \\
			Core-mantle-crust &	Core-mantle & 0.73 & 0.47 & 0.59 & 0.21 \\
			Core-mantle-crust & Uniform   & 4.81 & 2.38 & 3.12 & 0.91 \\
			& & &  & & \\
			PREM profile & Core-mantle-crust   & 0.36 & 0.24 & 0.30 & 0.11 \\
			PREM profile & Vacuum & 5.52 & 3.52 & 4.09 & 1.67 \\
			PREM profile & Mantle-crust & 7.45 & 3.76 & 4.83 & 1.59 \\
			PREM profile & Core-mantle & 0.27 & 0.18 & 0.21 & 0.07 \\
			PREM profile & Uniform  & 6.10 & 3.08 & 3.92 & 1.18 \\
			\hline \hline
		\end{tabular}
	\end{center}
	\mycaption{Ruling out the alternative profiles of Earth at the median $\Delta \chi^2$ level. We marginalize over oscillation parameters $\sin^2\theta_{23}$, $\Delta m^2_\text{eff}$, and mass ordering in theory, whereas remaining oscillation parameters are kept fixed at their benchmark values as mentioned in Table~\ref{tab:osc-param-value}.  
		The third and fourth (fifth and sixth) columns show results considering NO (IO) as true mass ordering in data. The results in the third and fifth (fourth and sixth) columns are with (without) the charge identification capability of ICAL.}
	\label{tab:chisq_analysis_results}
\end{table}

Till now, we have shown the fixed-parameter results, but now for final results, we marginalize over oscillation parameters $\sin^2\theta_{23}$, $\Delta m^2_\text{eff}$ and mass ordering while incorporating systematic errors as explained in Section~\ref{sec:statistical analysis}.
The total statistical significance includes contributions from both $\mu^-$ as well as $\mu^+$ as shown in Eq.~\ref{eq:chisq_total}. Here, we calculate the statistical significance to rule out the alternative profiles of Earth in theory with respect to the three-layered profile of core-mantle-crust in data as shown in Table~\ref{tab:chisq_analysis_results}. We have also compared alternative profiles of Earth in theory with respect to the PREM profile~\cite{Dziewonski:1981xy} in MC data. We would like to remind you that the PREM profile is with 25 layers as described in Section~\ref{sec:earth_model} by the solid black line in the right panel of Fig.~\ref{fig:three-layer-model}.

We can observe in Table~\ref{tab:chisq_analysis_results} that the $\Delta \chi^2_\text{ICAL-profile}$ for ruling out the vacuum in theory with respect to the three-layered profile of core-mantle-crust in data is 4.65 for NO (true) with CID, which shows that ICAL has good sensitivity towards the presence of matter effect.  In the absence of CID, this $\Delta \chi^2_\text{ICAL-profile}$ drops to 2.96, which shows that the capability of ICAL to distinguish $\mu^-$ and $\mu^+$ is crucial to observe the matter effect. For the case of IO (true), these numbers decrease further because, in this case, most of the contribution comes from $\mu^+$ that has lesser statistics due to a lower cross-section of antineutrinos compared to neutrinos.

Since we have found that ICAL can sense the presence of matter effect, now we can calculate the statistical significance to identify the profile that satisfies the distribution of matter inside Earth. The $\Delta \chi^2_\text{ICAL-profile}$ for ruling out the two-layered coreless profile of mantle-crust in theory with respect to the three-layered core-mantle-crust profile in the prospective data is about 6.31 for NO (true) with CID, and this is the sensitivity with which ICAL can validate the presence of core inside Earth. For the case of IO (true), this result drops to 3.92. 

We find that the trend in the final results with marginalization is the same as observed for the fixed-parameter case. After marginalization, the contributions from $\mu^-$ ($\mu^+$) events towards the $\Delta \chi^2_\text{ICAL-profile}$ for validating Earth's core is 6.09 (0.21) for NO as the true choice of mass ordering. If IO is the true mass ordering, then we see an opposite trend where the contribution towards the $\Delta \chi^2_\text{ICAL-profile}$ from $\mu^-$ ($\mu^+$) is 0.09 (3.82) after marginalization.

We would like to mention that if we do not incorporate hadron energy information and just use ($E_\mu^\text{rec}$, $\cos\theta_\mu^\text{rec}$) binning scheme from Table~\ref{tab:binning-2D-10years} then the $\Delta \chi^2_\text{ICAL-profile}$ for validating Earth's core after marginalization over oscillation parameters is about 3.20 for NO (true) with CID. Thus, we can say that the incorporation of hadron energy information improves the sensitivity of ICAL towards validating Earth's core.

The $\Delta \chi^2_\text{ICAL-profile}$ for ruling out the core-mantle profile in theory with respect to the core-mantle-crust profile in data is smaller than 1, which shows that the matter effect caused by crust is not significant. For ruling out the uniform distribution of matter in theory, we get  $\Delta \chi^2_\text{ICAL-profile}$ as 4.81, which indicates the capability of ICAL to feel the non-uniformity in density distribution inside Earth. 

We would like to mention that it does not make much difference if we perform analysis using the simple three-layered profile instead of the PREM profile and save computational time. The $\Delta \chi^2_\text{ICAL-profile}$ for the three-layered profile of core-mantle-crust in theory with respect to 25-layered PREM profile (as shown by the black line in the right panel of Fig.~\ref{fig:three-layer-model}) in data is 0.36 (0.30) for NO (IO) which shows that irrespective of the choice of the ordering of neutrino masses in nature, the analysis of atmospheric neutrino data with the simplified three-layered profile is a legitimate choice. Note that if we generate our prospective data with the more refined PREM profile having 25 layers and try to distinguish it from our hypothetical mantle-crust profile in theory, then we get a slightly increased $\Delta \chi^2_\text{ICAL-profile}$ of 7.45 for NO and 4.83 for IO.

\subsection{Impact of marginalization over various oscillation parameters}
\label{sec:results_margin_impact}
The sensitivity of ICAL to differentiate various density profiles of Earth may get deteriorated due to the uncertainties in neutrino oscillation parameters. To understand the impact of uncertainties of individual oscillation parameters on the sensitivity of ICAL to rule out an alternative profile of Earth while generating prospective data with the three-layered profile of core-mantle-crust, we marginalize over one oscillation parameter at a time in theory as shown in Table~\ref{tab:margin_impact_chisq}. In data, we take NO as true mass ordering and use benchmark values of oscillation parameter given in Table~\ref{tab:osc-param-value}. 

The third column of Table~\ref{tab:margin_impact_chisq} shows the fixed-parameter $\Delta \chi^2_\text{ICAL-profile}$ where we have not marginalized over any oscillation parameters in theory. In the fourth column, we marginalize over $\sin^2\theta_{23}$ in the range (0.36, 0.66) in theory and keep the other oscillation parameter fixed at their benchmark values as mentioned in Table~\ref{tab:osc-param-value}. Similarly, we marginalize over $|\Delta m^2_\text{eff}|$ in the range (2.1, 2.6) $\times 10^{-3} ~\text{eV}^2$ with same mass ordering (NO) in theory and data as shown in the fifth column. In the sixth column, we marginalize over $\Delta m^2_\text{eff}$ while considering both mass orderings in theory which effectively varies $\Delta m^2_\text{eff}$ in the range (-2.6, -2.1) $\times 10^{-3}~\text{eV}^2$ and (2.1, 2.6) $\times 10^{-3}~\text{eV}^2$. Finally, in last column, we shows $\Delta \chi^2_\text{ICAL-profile}$ with marginalization over $\sin^2\theta_{23}$, $\Delta m^2_\text{eff}$, and both mass orderings in theory.

\begin{table}[hbt!]
	\begin{center}
		\begin{tabular}{|l|l|c|c|c|c|c|}
			\hline \hline
			\multirow{3}{*}{MC Data} & \multirow{3}{*}{Theory} & \multicolumn{5}{c|}{$\Delta \chi^2_\text{ICAL-profile}$} \\ \cline{3-7}
			 &  & Fixed & \multicolumn{4}{c|}{Marginalization over}\\ \cline{4-7}
			&  & parameter & $\sin^2\theta_{23}$ & $|\Delta m^2_\text{eff}|$ & $\pm|\Delta m^2_\text{eff}|$ & All \\ \hline
			Core-mantle-crust &	Mantle-crust & 6.90 & 6.36 & 6.84 & 6.84 & 6.31 \\
			Core-mantle-crust &	Vacuum & 6.80 & 6.44 & 5.16 & 4.94 & 4.65 \\
			PREM &	Mantle-crust & 7.88 & 7.47 & 7.81 & 7.81 & 7.45 \\
			PREM &	Vacuum & 7.71 & 7.28 & 6.10 & 5.89 & 5.52 \\
			\hline \hline
		\end{tabular}
	\end{center} 

	\mycaption{The impact of marginalization over oscillation parameters $\sin^2\theta_{23}$, $|\Delta m^2_\text{eff}|$, and mass ordering on the sensitivity of ICAL to rule out the alternative profile of Earth at the median $\Delta \chi^2$ level. We assume true mass ordering as NO in data. The $\Delta \chi^2_\text{ICAL-profile}$ for the fixed-parameter case is given in the third column. The marginalized $\Delta \chi^2_\text{ICAL-profile}$ obtained after performing minimization separately over $\sin^2\theta_{23}$, $|\Delta m^2_\text{eff}|$, and $\Delta m^2_\text{eff}$ (with both mass orderings) in theory are given in fourth, fifth and sixth columns, respectively. The marginalized $\Delta \chi^2_\text{ICAL-profile}$ after performing combined minimization over $\sin^2\theta_{23}$, $\Delta m^2_\text{eff}$, and both mass orderings in theory is given in the last column. The remaining oscillation parameters are kept fixed at their benchmark values as mentioned in Table~\ref{tab:osc-param-value}.}
	\label{tab:margin_impact_chisq}
\end{table}
	
    The median $\Delta \chi^2_\text{ICAL-profile}$ which is the sensitivity of ICAL to rule out the two-layered profile of mantle-crust while generating prospective data with the three-layered profile of core-mantle-crust, is 6.90 when no marginalization is performed over any oscillation parameter as shown in the first row of Table~\ref{tab:margin_impact_chisq}. After marginalization over $\sin^2\theta_{23}$, $\Delta m^2_\text{eff}$, and both mass orderings in theory, the above-mentioned $\Delta \chi^2_\text{ICAL-profile}$ drops to 6.31. Here, marginalization over $\sin^2\theta_{23}$ in theory affects the sensitivity most.	
    
    Similarly, when we rule out vacuum scenario in theory by generating data with the core-mantle-crust profile, we obtain $\Delta \chi^2_\text{ICAL-profile}$ of 6.80 if we do not marginalize over any oscillation parameter in theory as shown in the second row of Table~\ref{tab:margin_impact_chisq}. This $\Delta \chi^2_\text{ICAL-profile}$ reduces to 4.65 if we marginalize over $\sin^2\theta_{23}$, $\Delta m^2_\text{eff}$, and both mass orderings in theory. We observe that in this case, the marginalization over $\Delta m^2_\text{eff}$, and both mass orderings substantially reduces the $\Delta \chi^2_\text{ICAL-profile}$. 
    
    From the above-mentioned observations, we can conclude that the marginalization over oscillation parameters has a large impact when we attempt to distinguish between various density profiles at ICAL. In the future, the more precise determination of oscillation parameters will help us to rule out the alternative profiles with better sensitivity at ICAL. The above findings hold if we generate the prospective data with the 25-layered PREM profile instead of the three-layered profile of core-mantle-crust and differentiate it against the mantle-crust or vacuum profile in theory. 
   
\subsection{Impact of different true choices of $\sin^2\theta_{23}$}
\label{sec:results_th23}

\begin{figure}
	\centering
	\includegraphics[width=0.49\linewidth]{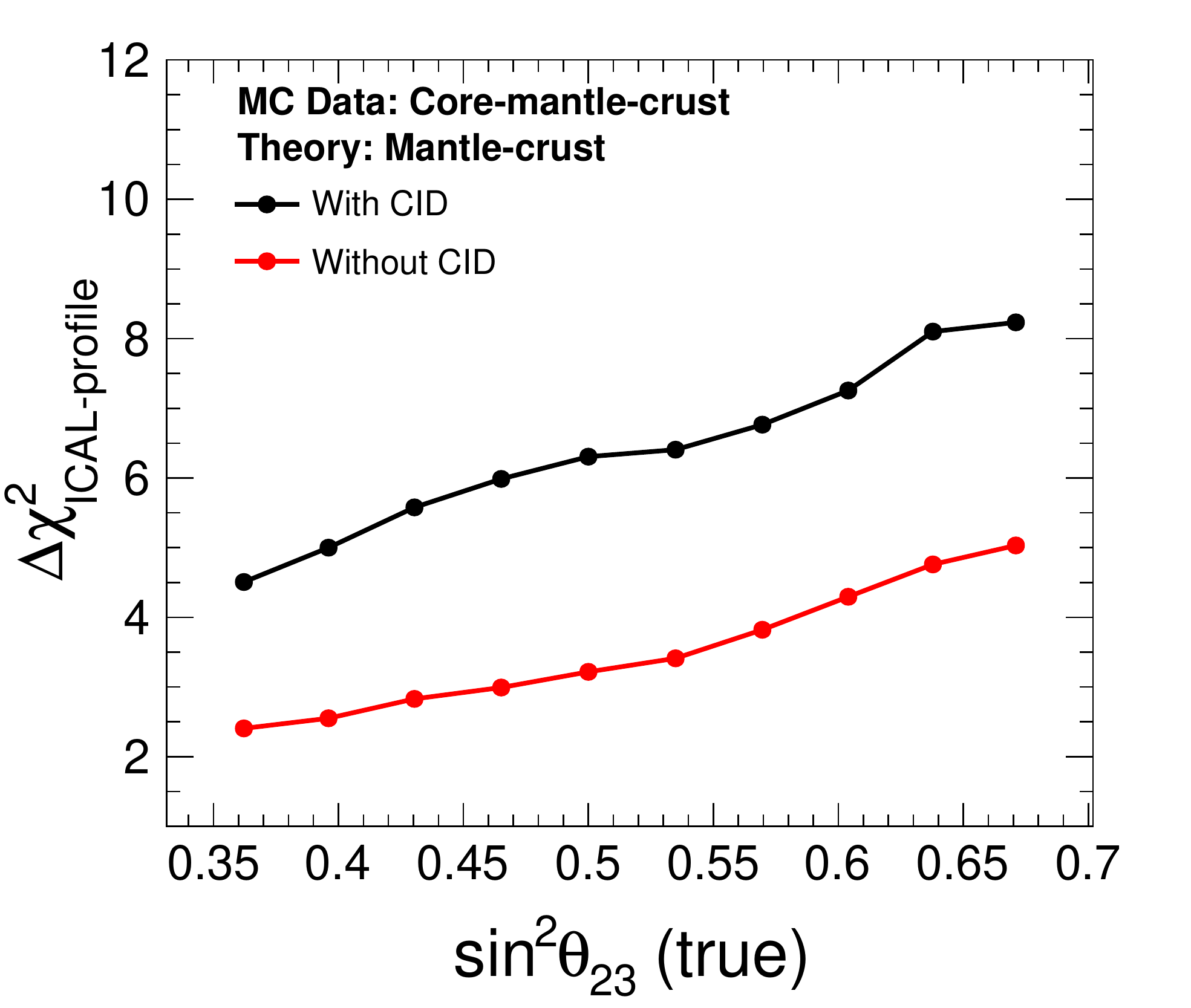}	\includegraphics[width=0.49\linewidth]{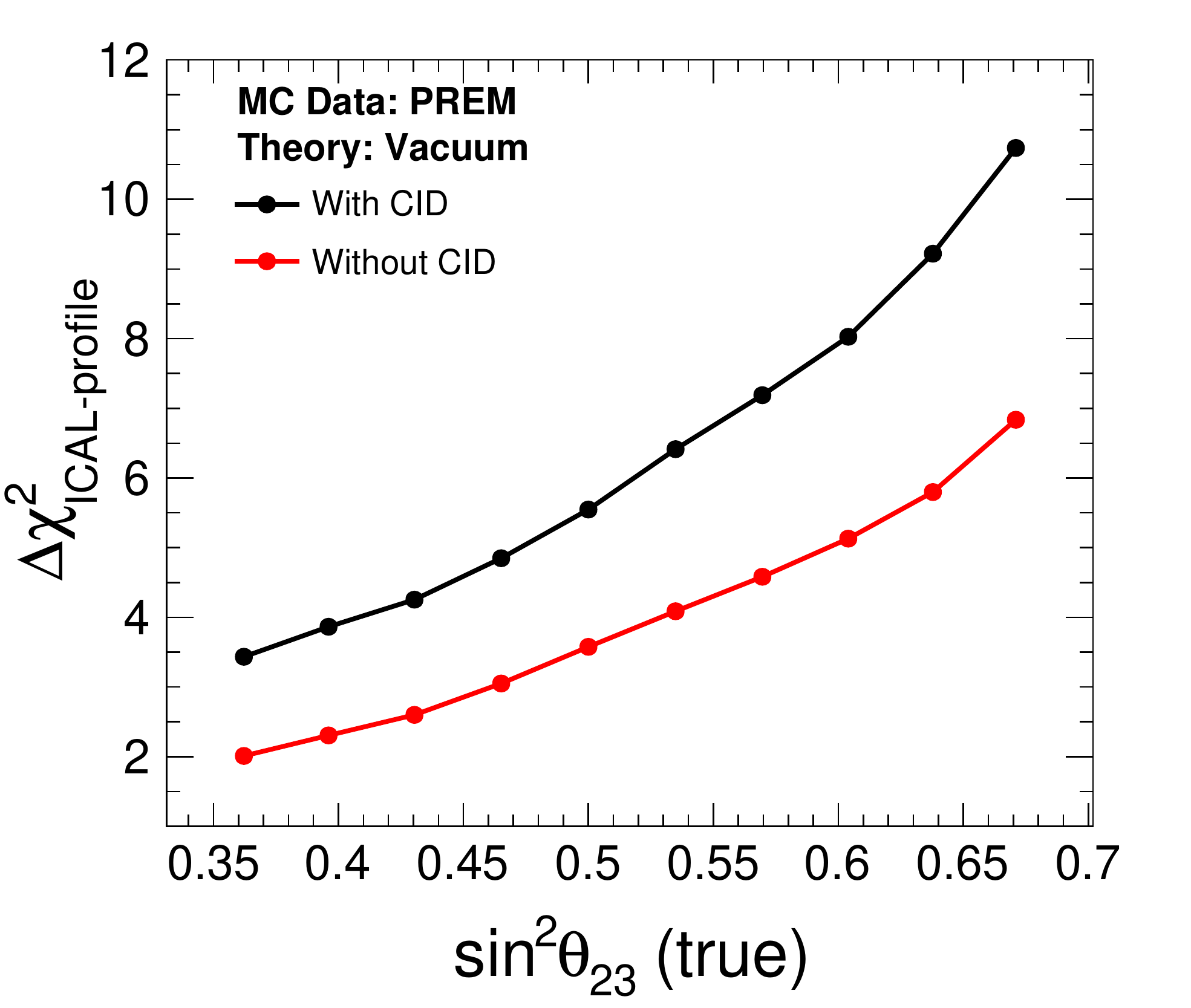}
	\mycaption{The median $\Delta \chi^2_\text{ICAL-profile}$ as a function of the choice of $\sin^2\theta_{23}$ in data. 
	The median $\Delta \chi^2_\text{ICAL-profile}$ is the sensitivity with which we can validate Earth's core by ruling out 
	the mantle-crust profile in theory w.r.t. the core-mantle-crust profile in data as shown in the left panel. The right panel 
	shows the sensitivity at median $\Delta \chi^2_\text{ICAL-profile}$ level to rule out vacuum scenario in theory w.r.t. 
	the PREM profile in data. In both the panels, the black (red) curve shows the sensitivity with (without) charged 
	identification capability of ICAL. Note that we marginalize over oscillation parameters $\sin^2\theta_{23}$, 
	$\Delta m^2_\text{eff}$, and mass ordering, whereas the remaining oscillation parameters are kept fixed 
	at their benchmark values as mentioned in Table~\ref{tab:osc-param-value}. We assume mass ordering 
	as NO in data.}
	\label{fig:chisq_th23_var}
\end{figure}

So far, we have taken in our analysis, $\sin^2\theta_{23}(\text{true}) = 0.5$ as our benchmark choice but the recent global fit data also indicates that $\theta_{23}$ may not be maximal, it can either lie in the lower octant where $\sin^2\theta_{23} < 0.5$ or the higher octant where $\sin^2\theta_{23} > 0.5$. Needless to mention that $\theta_{23}$ is the most uncertain oscillation parameter at present apart from $\delta_{\rm CP}$. So, now, it is legitimate to see how the sensitivity of ICAL towards validating the Earth's core may change if, in nature, $\theta_{23}$ (true) turns out to be non-maximal. To analyze this, we are presenting Fig.~\ref{fig:chisq_th23_var} where, in the x-axis, we have varied the choice of $\sin^2 \theta_{23}$ in data in the range 0.36 to 0.66, and in the y-axis, we are evaluating the median $\Delta \chi^2_\text{ICAL-profile}$, the sensitivity with which we can validate Earth' core (left panel) and rule out vacuum scenario in theory with respect to PREM profile in data (right panel). Here, we marginalize over oscillation parameters $\sin^2\theta_{23}$ in the range of 0.25 to 0.75, $\Delta m^2_\text{eff}$ in the range of (2.1, 2.6) $\times 10^{-3}~\text{eV}^2$ and both the mass orderings NO as well as IO, whereas the remaining oscillation parameters are kept fixed at their benchmark values as mentioned in Table~\ref{tab:osc-param-value}. 

The dominant contribution of matter effect appears in term of $\sin^2\theta_{23}$ for survival probability $P(\nu_\mu\rightarrow \nu_\mu)$ as well as appearance probability $P(\nu_e\rightarrow \nu_\mu)$ as shown by series expansion in Ref.~\cite{Akhmedov:2004ny}. $P(\nu_\mu\rightarrow \nu_\mu)$ decreases almost linearly with $\sin^2\theta_{23}$ whereas $P(\nu_e\rightarrow \nu_\mu)$ increases linearly. Since the contribution of appearance $(\nu_e\rightarrow \nu_\mu)$ channel is smaller than that of survival $(\nu_\mu\rightarrow \nu_\mu)$ channel, the net matter effect do not cancel out completely and show almost linear dependence on $\sin^2\theta_{23}$.

This linear dependence of matter effect on $\sin^2\theta_{23}$ results in an increasing $\Delta \chi^2_\text{ICAL-profile}$ with $\sin^2\theta_{23}$(true) as shown in both panels in Fig.~\ref{fig:chisq_th23_var} because $\Delta \chi^2_\text{ICAL-profile}$ in both the cases are driven by matter effect. Thus, we can say that the Earth's core can be validated with a higher confidence level if, in nature, $\theta_{23}$ is found to be lying in the higher octant. We can also observe in both cases that $\Delta \chi^2_\text{ICAL-profile}$ is higher if the charge identification capability is present.  Thus, the presence of charge identification capability is crucial in validating Earth's core (left panel) as well as ruling out vacuum scenario (right panel).

\section{Summary and concluding remarks}
\label{sec:conclusion}

Atmospheric neutrinos travel long distances inside Earth and feel the presence of matter effect that depends upon the density distribution inside Earth. Neutrino oscillation tomography utilizes the matter effect experienced by neutrinos to unravel the internal structure of Earth. Guided by the PREM profile, we use a three-layered density profile of Earth where we have core, mantle, and crust. For comparison, we consider alternative profiles of Earth -- mantle-crust, core-mantle, and uniform density. 

In Section~\ref{sec:probability}, we show the effect for various profiles of Earth on the neutrino oscillations in $P(\nu_\mu\rightarrow\nu_\mu)$ and $P(\nu_e\rightarrow\nu_\mu)$ channels. We observe that the presence of mantle and core result in the MSW resonance and NOLR/parametric resonance, respectively. On the other hand, the presence of a boundary between layers results in a sharp transition in oscillation probabilities in $P(\nu_e\rightarrow\nu_\mu)$ channel.

Table~\ref{tab:events} shows that about 4614 $\mu^-$ and 2053 $\mu^+$ events are expected at ICAL for 500 kt$\cdot$yr exposure considering three-flavor neutrino oscillations in the presence of matter with PREM profile. Utilizing the neutrino direction, we estimate that about 331 $\mu^-$ and 146 $\mu^+$ core-passing events would be detected at ICAL in 10 years. The events passing through the crust-mantle-core region and only crust are shown in Table~\ref{tab:layer-passing-event}. In Fig.~\ref{fig:event_dist_nu_density_zones}, we can observe that the information about the region traversed by neutrinos is preserved even after reconstruction as muon events, but some of the reconstructed muons may get smeared into other regions due to reaction kinematics and finite detector resolution.

After identifying the events passing through various regions inside Earth, we perform statistical analysis to differentiate between two profiles of Earth using atmospheric neutrino events at ICAL. We would like to mention that $\Delta \chi^2$ for the determination of mass ordering is contributed by both neutrino and antineutrino irrespective of the choice of true mass ordering. On the other hand, in our study where we are contrasting between different profiles of Earth for a given mass ordering, $\Delta \chi^2$ is mostly contributed by neutrino for NO (true) and antineutrino for IO (true). We estimate statistical significance at $\Delta \chi^2$ level to rule out the coreless profile of mantle-crust with respect to core-mantle-crust profile as given by Eq.~\ref{eq:chisq_diff}. Figure~\ref{fig:chisq_contour} shows that the significant contribution to $\Delta \chi^2_{-}$ (NO) and $\Delta \chi^2_+$ (IO) is received from higher baselines and lower energies which is the region around the boundary between core and mantle. The density in this region gets significantly modified in the absence of a core.  

We show the final results in Table~\ref{tab:chisq_analysis_results} in terms of $\Delta \chi^2_\text{ICAL-profile}$ to rule out the alternative profiles in theory with respect to core-mantle-crust profile in data. For final results, $\Delta \chi^2_\text{ICAL-profile}$ is marginalized over oscillation parameters $\sin^2\theta_{23}$, $\Delta m^2_\text{eff}$ and mass ordering. The results for the coreless profile of mantle-crust in theory with respect to the core-mantle-crust profile in prospective data show that the presence of Earth's core can be validated at $\Delta \chi^2_\text{ICAL-profile}$ of 6.31 for NO (true) and 3.92 for IO (true) using 500 kt$\cdot$yr exposure at ICAL with charge identification capability. On the other hand, if we generate our prospective data with a more refined PREM profile of the Earth having 25 layers and contrast it with our hypothetical profile of the Earth consisting of only mantle and crust in theory, then we get a slightly enhanced $\Delta \chi^2_\text{ICAL-profile}$ of 7.45 for NO (true) and 4.83 for IO (true). Important to note that in the absence of charge identification capability of ICAL, these sensitivities deteriorate significantly to 3.76 for NO (true) and 1.59 for IO (true).

We demonstrate that the sensitivity to rule out the alternative profiles of Earth deteriorates with marginalization. This indicates that with the improvement in the precision of oscillation parameters in the future, the alternate profiles of Earth can be ruled out with better sensitivity. In Fig.~\ref{fig:chisq_th23_var}, we show that the sensitivity to validate Earth's core increases as we increase the true value of $\sin^2\theta_{23}$. Thus, the presence of Earth's core can be validated at higher sensitivity if $\theta_{23}$ is found to be lying in the higher octant. It is important to note that the presence of charge identification capability is an important feature of ICAL, which significantly improves the results for studies involving matter effect. We hope that the analysis performed in this paper will open a new vista for the ICAL detector at the upcoming INO facility.

\subsubsection*{Acknowledgements}

This work is performed by the members of the INO-ICAL collaboration to explore the possibility of utilizing neutrino oscillations in the presence of matter to extract information about the internal structure of Earth complementary to the seismic studies. We thank A. Dighe, M. V. N. Murthy, S. Petcov, V. M. Datar, N. K. Mondal, S. Uma Sankar, A. Smirnov, F. Halzen, P. Coyle, E. Lisi, S. Palomares-Ruiz, S. Goswami, D. Indumathi, P. Roy, and S. P. Behera, for their useful suggestions and constructive comments on our work. A. Kumar would like to thank the organizers of the XXIV DAE-BRNS High Energy Physics Online Symposium at NISER, Bhubaneswar, India, during 14th to 18th December 2020, for providing him an opportunity to present a poster based on this work. We acknowledge the support of the Department of Atomic Energy (DAE), Govt. of India. S.K.A. is supported by the DST/INSPIRE Research Grant [IFA-PH-12] from the Department of Science and Technology (DST), Govt. of India, and the Young Scientist Project [INSA/SP/YSP/144/2017/1578] from the Indian National Science Academy (INSA). S.K.A. acknowledges the financial support from the Swarnajayanti Fellowship Research Grant (No. DST/SJF/PSA-05/2019-20) provided by the Department of Science and Technology (DST), Govt. of India and the Research Grant (File no. SB/SJF/2020-21/21) provided by the Science and Engineering Research Board (SERB) under the Swarnajayanti Fellowship by the DST, Govt. of India. The numerical simulations are performed using SAMKHYA: High-Performance Computing Facility at Institute of Physics, Bhubaneswar.

\bibliographystyle{JHEP}
\bibliography{Tomography-References.bib}

\end{document}